\documentclass[]{hdsr}

\usepackage{moreverb,url}
\usepackage{pdflscape}
\usepackage{soul}

\usepackage[colorlinks,bookmarksopen,bookmarksnumbered,citecolor=red,urlcolor=red]{hyperref} 
\graphicspath{{figs/}}

\newcommand\BibTeX{{\rmfamily B\kern-.05em \textsc{i\kern-.025em b}\kern-.08em
T\kern-.1667em\lower.7ex\hbox{E}\kern-.125emX}}

\begin{document}

\begin{center}
\Large A comparison of mixed-models for the analysis of non-linear longitudinal data: application to late-life cognitive trajectories
\end{center}

\vspace{5mm}

  \thispagestyle{empty}

  \begin{tabular}{cc}
  Maude Wagner\upstairs{\affilone,*}, 
    Donald R. Hedeker\upstairs{\affiltwo}, 
    Tianhao Wang\upstairs{\affilone}, \\
    Graciela Muniz-Terrera\upstairs{\affilthree},
    Ana W. Capuano\upstairs{\affilone}
   \\
   {\small \upstairs{\affilone} Rush Alzheimer’s Disease Center, Rush University Medical Center, Chicago, IL, USA} \\
   {\small \upstairs{\affiltwo} Department of Public Health Sciences, The University of Chicago Biological Sciences, Chicago, IL, USA} \\
   {\small \upstairs{\affilthree}Edinburgh Dementia Prevention Group, University of Edinburgh, Edinburgh, United Kingdom} 
  \\
  \upstairs{*}Corresponding author: maude\_wagner@rush.edu
     
  \end{tabular}
  \\
  
  \emails{}

\normalsize{ABSTRACT. Several mixed-effects models for longitudinal data have been proposed to accommodate the non-linearity of late-life cognitive trajectories and assess the putative influence of covariates on it. No prior research provides a side-by-side examination of these models to offer guidance on their proper application and interpretation. In this work, we examined five statistical approaches previously used to answer research questions related to non-linear changes in cognitive aging: the linear mixed model (LMM) with a quadratic term, LMM with splines, the functional mixed model, the piecewise linear mixed model, and the sigmoidal mixed model. We first theoretically describe the models. Next, using data from two prospective cohorts with annual cognitive testing, we compared the interpretation of the models by investigating associations of education on cognitive change before death. Lastly, we performed a simulation study to empirically evaluate the models and provide practical recommendations. Except for the LMM-quadratic, the fit of all models was generally adequate to capture non-linearity of cognitive change and models were relatively robust. Although spline-based models have no interpretable nonlinearity parameters, their convergence was easier to achieve, and they allow graphical interpretation. In contrast, piecewise and sigmoidal models, with interpretable non-linear parameters, may require more data to achieve convergence.}

\section{INTRODUCTION}
Longitudinal characterization of cognitive change in late life has attracted the attention of numerous researchers to better understand normative cognitive aging and cognitive decline reflecting pathology-related and mortality-related processes [1,2,3,4]. Given the heterogeneity in cognitive paths observed in longitudinal studies of cognitive aging, mixed models of fixed and random effects have become a tool of choice to estimate late-life marginal and individual cognitive trajectories and assess the putative influence of covariates on it [5]. They offer practical advantages, including the treatment of intra-subject correlations, and the ability to handle intermittent missingness and varying times and numbers of observations per participant. Their applications have provided valuable insights into the natural history of specific conditions over time, such as dementia [4,6] and death [2].
\\
Late-life cognitive change can be markedly non-linear, with prior research reporting, for example, an accelerated decline starting approximatly 3 to 7 years before death [3,7]. Several statistical approaches have been proposed to accommodate the non-linearity of late-life cognitive trajectories in various research settings. Distinct in their structures, computational needs, and interpretations, these approaches can be particularly useful to provide complementary insights in characterizing cognitive decline and estimating the putative effects of covariates on it. For example, the piecewise linear mixed model is an appealing approach to detect the changepoint time when the onset of accelerated cognitive decline occurs [16,17,18]. The sigmoidal mixed model also offers great flexibility in fitting nonlinear trajectories and allows meaningful parametric quantities to be examined, such as early levels, late levels and half-decline [19,20]. However, no prior work provides a side-by-side examination of these models to offer guidance on their proper application and interpretation. Most importantly, no prior research provides a side-by-side simulation study of these models to offer a fair comparison of empirical properties.
\\
A review of all existing statistical approaches capturing non-linear trends in longitudinal outcomes is beyond the scope of this study. We propose here to examine the five mixed models previously used to answer specific research questions related to non-linear changes in cognitive aging, including the linear mixed model (LMM) with a quadratic term [9,10,11], the LMM with the function of time approximated directly from the data by a flexible basis of splines [12,13], the functional mixed model (also termed varying-coefficient model) [14,15], and two non-linear mixed models, including the piecewise linear mixed model with a random changepoint [16,17,18] and the sigmoidal mixed model with four parameters [19,20]. 
\\
The present manuscript is organized as follows. In Section 2, we provide a brief introduction of the aforementioned statistical models and comment on how each of them can help answer answering specific research questions related to cognitive aging. In Section 3, using real data from 1276 older deceased participants with $\geq$4 annual cognitive tests, we compare the interpretations of the models by analyzing the association of education on cognitive decline up to two decades before death. In Section 4, we employed a simulation study to challenge all the models and provide practical recommendations for model choices and sample requirements. In particular, we examined different patterns of cognitive change, missingness, and follow-up intervals. Section 5 concludes with a discussion and guidance. 
\\
\section{Statistical mixed models for longitudinal data}
\subsection{General setting}
\begin{figure}[h]
\centerline{\includegraphics[width=220pt]{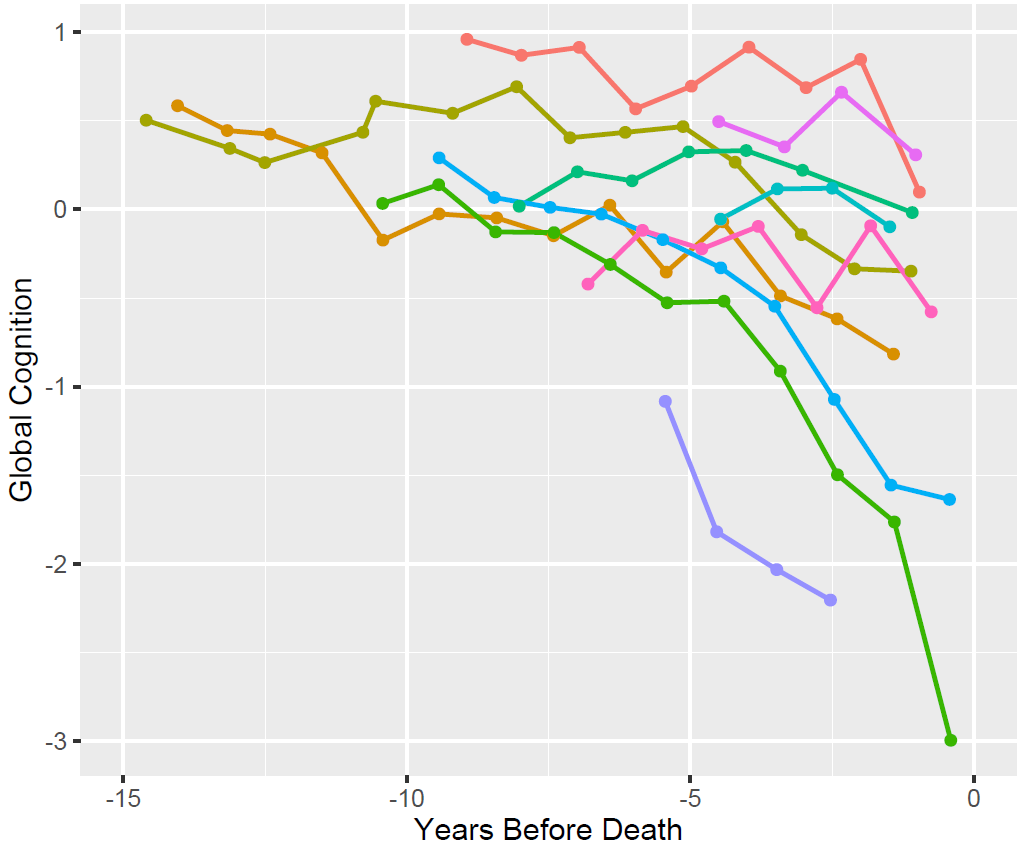}}
\caption{Theoretical examples of cognitive trajectories observed before death for 10 individuals with at least 4 repeated cognitive measures. }\label{fig1}
\end{figure}

Initially motivated by the longitudinal characterization of cognitive change in late life, all statistical mixed models considered in this work will be described within the following study framework. For each subject $i$ ($i$ = 1,..., $N$), consider a vector of $n_i$ repeated measures $y_i$ = ($y_{i1}$,..., $y_{ij}$,..., $y_{in_i})^\top$ where $y_{ij}$ denotes global cognition measured at the retrospective observation time $t_{ij}$ ($j$ = 1, …, $n_i$) before death, so that $t_{ij} \in [-T, 0]$. Note that measurement times and the total number of measures $n_i$ can differ from one individual to another (Figure \ref{fig1}). 

\noindent In the following sub-sections, we briefly introduce the five mixed models of interest, including LMMs with a quadratic term or with the function of time approximated directly from the data using splines (Section 2.2.), the functional mixed model (Section 2.3.), the piecewise linear mixed model with a random changepoint (Section 2.4), and the sigmoidal mixed model with four parameters (Section 2.5.). As we are considering a relatively large number of models, for simplicity, some annotations can be similar from one model to another, but the interpretation of the parameters remains specific to each one.  
\\
\subsection{Linear mixed models}
Longitudinal cognitive trajectories can be modeled using the classic LMM [8] formulation:
\vspace{3mm}
\begin{equation}\label{1}
y_{ij} = \psi_{0i} + F(t_{ij})^\top\psi_{1i} + \epsilon_{ij} 
\end{equation}

\begin{center} 
with  
\end{center}

\begin{center}
$\psi_{0i} = \alpha_0 + X_{0i}^\top \beta_{0} + \eta_{0i}$, 
\vspace{3mm}
\\
$\psi_{1i} = \alpha_1 + X_{1i}^\top \beta_{1}  + \eta_{1i}$,
\vspace{3mm}
\\
$\eta_i = (\eta_{0i}, \eta_{1i})^\top \sim \mathcal{N}(0, B)$ with $B$ = $\Bigl($ 
$\begin{matrix}
\sigma_0 & \sigma_{01}
\\ 
\sigma_{01} & \sigma_1
\end{matrix}$
$\Bigl)$ ~ and ~ $\epsilon_{ij} \sim \mathcal{N}(0,\sigma_{\epsilon})$
\end{center}

\noindent In this model \eqref{1}, $F(t)$ represents the function of time, which can be of any type (e.g., linear, splines); $\psi_{0i}$ and $\psi_{1i}$ represent the person-specific level at death ($t=0$) and person-specific slope before death, respectively; $\alpha_0$ and $\alpha_1$ represent the mean level at death and mean slope before death, respectively; $X_{0i}$ and $X_{1i}$ are vectors of explanatory variables associated with the vector of fixed effects $\beta_{0}$ and $\beta_{1}$, respectively; $\eta_{0i}$ and $\eta_{1i}$ are correlated person-specific random intercept and random slope, respectively; and $\epsilon_{ij}$ are random errors.
\\
\noindent In this work, we considered  two specific shapes of trajectories commonly used in the context of non-linear late-life cognitive decline. The first LMM assumes a quadratic change over time ($F(t) = t, t^{2}$; model termed hereafter LMM-quadratic). LMM-quadratic represents one of the oldest mixed models utilized in the realm of longitudinal data analysis. However, when the observed trajectory does not conform to quadratic structures, the use of semi-parametric trajectory modeling may be preferable. Thus, the second LMM considers a flexible function of time approximated directly from the data using splines [21,22], which allows for a nonlinear change in time without a priori assumption on the shapes of trajectories (model termed hereafter LMM-splines). For LMM-splines, $F(t)$ is replaced by $\sum_{k=1}^K \theta_{k}B_{k}(t)$ where $(B_{k})_{k=1,...,K}$ refers to the $K$ basis of spline functions and $(\theta_{k})_{k=1,...,K}$ the coefficients to estimate. Although any type of splines could be considered, in this work we examine the natural cubic splines that limit erratic behavior at the boundary of the observation time period thanks to linearity constraints [23]. As a practical rule, the number of inner nodes should be less than or equal to the average number of repeated measures. A large number of internal K-nodes implies high flexibility but can also lead to overfitting the data. Conversely, a small number of knots can lead to an over smooth estimate. Typically, the choice of the final spline basis relies on the right balance between the number of knots and the goodness of fit of the model assessed via parsimony and the Bayesian Information Criterion (BIC), where smaller values are preferable [24]. This choice can also be facilitated by inspecting the gain in flexibility of the marginal trajectories estimated using different numbers of nodes.

\subsection{The functional mixed model}

The FMM [14,15] extends the traditional LMM models by allowing to investigate the functional (non-linear) effects of any given predictor using nonparametric curves. It incorporates fixed and random effects modeled as smoothing splines that allows for flexibility but generates non-interpretable parameters. 
The FMM can be specified as follows:

\begin{equation}\label{8}
y_{ij} = \psi(t_{ij})^\top X_i + \gamma_{i}(t_{ij}) + \epsilon_{ij} 
\end{equation}

\noindent where $X_i$ is a vector of covariates associated with a $p$~$\times$ 1 vector of fixed functions $\psi(t) = (\psi_1(t),..., \psi_p(t))^\top$; $\gamma_i(t)$ is the person-specific random function; and $\epsilon_{ij}$ is the measurement error. The fixed functions $\psi(t)$ can be interpreted as the longitudinal marginal average profiles and $\gamma_i(t)$ captures the $i^{th}$ person-specific deviation from the marginal curve. Thus, $\psi(t)^\top X_i + \gamma_i(t)$ is the $i^{th}$ person-specific functional curve. Note that when $\psi(t)$ and $\gamma_i(t)$ are modeled as parametric functions, the FMM model in Equation \eqref{8} reduces to a standard LMM model.
\\
Conceptually, any smoothing method can be used to model the unknown coefficient functions $\psi(t)$ and $\gamma_i(t)$. Following Rice and Wu [14], we used the smoothing B-splines to model both $\psi(t)$ and $\gamma_i(t)$, such that:
\begin{equation}\label{9}
\begin{split}
\psi(t) &= B(t)^\top \phi_\psi
\\ \gamma_{i}(t) &= B(t)^\top \phi_{\gamma_i}
\end{split}
\end{equation}

\noindent where $B(t)$ refers to a $Z\times$1 vector of cubic B-spline basis functions with $Z$ – 4 inner knots defined in a fixed interval $\tau$ that contains all $t_i$ observations; $\phi_\psi$ and $\phi_{\gamma_i}$ denote the vectors of parameters for the fixed and the random functions, respectively. In our computations, we used equally spaced knots and the number of inner knots was based on the average total number of repeated measurements of the outcome, as previously recommended [25]. For the random coefficients, they are assumed to be homogeneous with the same Gaussian distribution, $\phi_{\gamma_i} \sim N(\phi_{\gamma}, \sum)$, where $\sum$ is the covariance matrix. Therefore, the mean function and the covariance kernel of the Gaussian process can be approximated by

\begin{equation}\label{10}
\mu(t) = B(t)^\top \phi_\mu
\end{equation}

\noindent with $\mu(t)$ representing the average cognitive profile. 
\\
Equations \eqref{9} and \eqref{10} use original B-spline expansions, which imply the use of a large number of basis functions to capture nonlinear patterns and a risk  of overfitting the fixed function $\psi$ and the mean function $\mu$. To overcome this issue, we used a penalized spline approach [26], and we transformed the B-splines basis functions into a linear part and a nonlinear part, in a similar fashion to the smoothing splines approach [15]. 
\\
Define $W = \int_{\tau}^{} B''(t)B''(t)^\top \,dt$. The rank of $W$ is $Z - 2$ so that the eigendecomposition of $W$ has a form as
\begin{equation}\label{11A}
\begin{split}
W = [V_1, V_2] \times	\begin{bmatrix} 
	  0_{2\times2} &  \\
	            & \Delta \\
	\end{bmatrix} \times [V_1, V_2]^\top
\end{split}
\end{equation}

\noindent where $V_1 = (v_{10},v_{11})$ is a $Z \times 2$ matrix; $V_2$ is a $Z\times(Z-2)$ matrix orthogonal to $V_1$; $\Delta$ is a diagonal matrix of $Z$ – 2 strictly positive numbers that contains the eigenvalues of $W$. Any two functions $B(t)^\top v_{10}$ and $B(t)^\top v_{11}$ that span the space of linear functions on $\tau$ can be orthogonalized to provide a valide $V_1$. Thus, the functional space spanned by $B(t)^\top v_{10}$ and $B(t)^\top v_{11}$ is equivalent to that of a constant function 1 and $t$. Then, the transformed B-spline basis functions used in Equations \eqref{9} and  \eqref{10} can be defined as 
\begin{equation}\label{11}
\begin{split}
\tilde{B}(t)^\top &= \left(1, t, \underbrace{{B}(t)^\top V_2 \Delta^{-1/2}}\right)^\top  
\\&= \left(1, t,  \hspace*{0.7cm} \tilde{B}_2(t)^\top \hspace*{0.6cm} \right)^\top 
\end{split}
\end{equation}

\noindent The fixed function $\psi(t)$ and the random function $\gamma_i(t)$ defined in Equation  \eqref{9} become
\\
\begin{equation}\label{12}
\begin{split}
\psi(t) &= b_0 + b_1t + \tilde{B}_2(t)^\top b_2
\\ \gamma_{i}(t) &= a_{0i} + a_{1i}t + \tilde{B}_2(t)^\top a_{2i}
\end{split}
\end{equation}
\\
And the mean function $\mu(t)$ defined in Equation \eqref{10} become
\begin{equation}\label{13A}
\begin{split}
\mu(t) &= u_0 + u_1t + \tilde{B}_2(t)^\top u_2
\end{split}
\end{equation}

\noindent $\psi(t)$, $\gamma_i(t)$, and $\mu(t)$ are divided into a linear component (i.e., 1, $t$) and a non-linear component (i.e., $\tilde{B}_2(t)$), and the transformed basis parameters to be estimated correspond to $\phi_\psi$ = ($b_0, b_1, b_2$), $\phi_{\gamma_i}$ = ($a_{0i}, a_{1i}, a_{2i}$), and $\phi_\mu = (u_0, u_1, u_2)$, respectively. 
\\
The LMM-quadratic, LMM-splines, and FMM are linear in the parameters and describe how the longitudinal outcome varies with covariates. These models do not include any theoretical considerations about the underlying mechanism producing the longitudinal data. In contrast, non-linear mixed models are non-linear in the parameters and allow for the inclusion of covariates related to meaningful parametric quantities.

\subsection{The piecewise linear mixed model with a random changepoint}
The PMM model [27,28] is a non-linear mixed model that assumes that the stochastic process of the longitudinal outcome is characterized by two or more different phases. In the context of our motivating application, late-life cognitive decline, a single changepoint, and two linear phases are commonly used [28,29,30]. In particular, this PMM model with an abrupt change (also known as the linear-linear or the broken-stick mixed model) provides an appealing statistical approach to detect the changepoint time when the onset of accelerated cognitive decline occurs. Although the PMM-abrupt consists of two linear phases intersecting at the changepoint, it is important to note that it remains nonlinear in its parameters. 
\\
A commonly used parameterization of the PMM-abrupt model [27] is: 

\begin{equation}\label{2}
\begin{split}
\mathbb{E}[y_{ij}] = 
\begin{cases}
\psi_{0i} + \psi_{1i}t_{ij} ~~~~~~~ &\mbox{ if   } t_{ij} < \psi_{3i}\\
\psi_{0i} + \psi_{1i}\psi_{3i} + \psi_{2i}(t_{ij} - \psi_{3i}) ~~ &\mbox{ if   } t_{ij} \geq \psi_{3i}
\end{cases}
\end{split}
\end{equation}

\begin{center} 
with  
\end{center}

\begin{center}
$\psi_{0i} = \alpha_0 + X_{0i}^\top\beta_{0} + \eta_{0i}$, 
\vspace{3mm}
\\
$\psi_{1i} = \alpha_1 + X_{1i}^\top\beta_{1} + \eta_{1i}$,
\vspace{3mm}
\\
$\psi_{2i} = \alpha_2 + X_{2i}^\top\beta_{2} + \eta_{2i}$,
\vspace{3mm}
\\
$\psi_{3i} = \alpha_3 + X_{3i}^\top\beta_{3} + \eta_{3i}$,
\vspace{3mm}
\\
$\eta_i =  (\eta_{0i}, \eta_{1i}, \eta_{2i}, \eta_{3i})^\top \sim \mathcal{N}(0, B)$ 
\end{center}

\begin{figure}[h]
\centerline{\includegraphics[width=270pt]{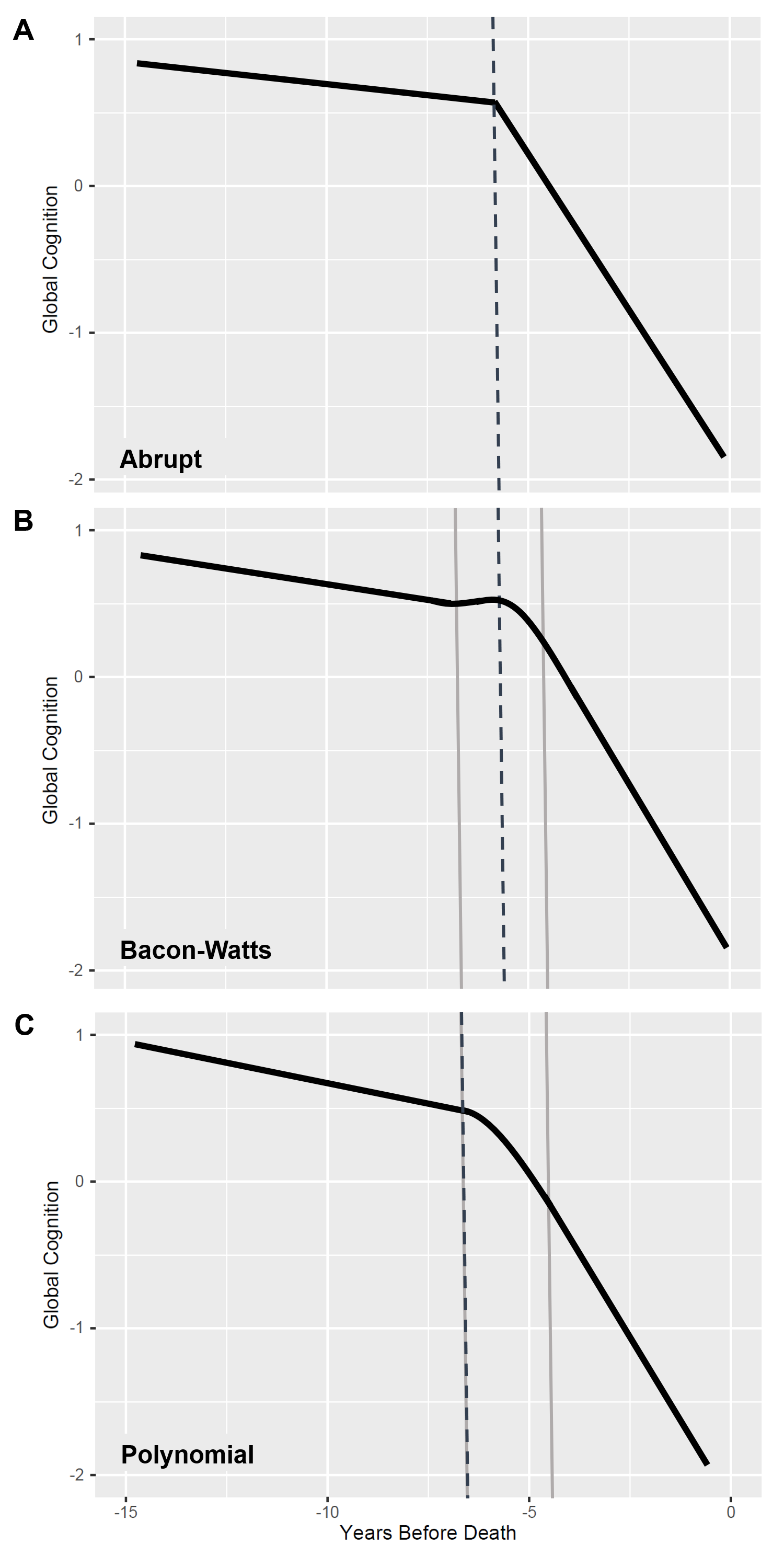}}
\caption{Example of trajectories based on \textbf{(A)} a piecewise linear mixed model with an abrupt change, \textbf{(B)} a Bacon-Watts model with an hyperbolic tangent transition, and \textbf{(C)} a piecewise linear mixed model with a polynomial smooth transition. The dashed black line indicates the changepoint time parameter estimated in the model. The solid gray lines represent the limits of the transition period. Adapted from Van den Hout and colleagues [18] and Segalas and colleagues [32]. }\label{fig2}
\end{figure}

\noindent In this model \eqref{2}, $\psi_{3i}$ represents the person-specific changepoint time where the abrupt change occurs; $\psi_{1i}$ and $\psi_{2i}$ represent the person-specific slopes before and after $\psi_{3i}$, respectively; $\psi_{0i}$ represents the person-specific level at the intersection of the first linear phase with the y-axis at time zero (intercept not directly interpretable; see Figure \ref{fig2}A); $\alpha_0$, $\alpha_1$, $\alpha_2$, and $\alpha_3$ represent the mean values for the intercept, the slope before the changepoint, the slope after the changepoint, and the changepoint time, respectively; $X_{0i}$, $X_{1i}$, $X_{2i}$, and $X_{3i}$ are vectors of covariates associated with the vector of fixed effects $\beta_{0}$, $\beta_{1}$, $\beta_{2}$, and $\beta_{3}$, respectively; and $\eta_{0i}$ to $\eta_{3i}$ are person-specific random effects with $(\eta_{0i}, \eta_{1i}, \eta_{2i}, \eta_{3i})^\top \sim MVN(0, B)$ and $B$ assuming correlations only between $\eta_{1i}$ and $\eta_{2i}$. 
\\
The PMM-abrupt defined in Equation \eqref{2} results in an abrupt change between the two phases (see Figure \ref{fig2}A). However, in the context of cognitive decline, given that neurodegeneration is a continuous process, a gradual shape of the change is more realistic. To accommodate a smooth transition between the two periods, several alternative approaches have been proposed, including the Bacon-Watts model with a hyperbolic tangent transition function [31] (see Figure \ref{fig2}B) and the PMM with a cubic polynomial transition function [18] (see Figure \ref{fig2}C). In the present work, we chose to focus on the PMM-polynomial. Indeed, compared with the Bacon-Watts model, for example, the PMM-polynomial presents three major advantages: (i) it allows an intuitive and direct interpretation of the parameters from the two linear phases, (ii) the changepoint parameter is defined as the onset of the transition phase (rather than the middle of the transition period for the Bacon-Watts model), and (iii) it uses a smooth polynomial transition function fitting a flexible monotonic curve between the two phases, while the Bacon-Watts model can produce an unrealistic bump over the transition [18,31].
The PMM-polynomial can be written as follows: 
\begin{equation}\label{3}
\begin{split}
\mathbb{E}[y_{ij}] = 
\begin{cases}
\psi_{0i} + \psi_{1i}t_{ij} ~~~~~~~ &\mbox{ if   } t_{ij} < \tilde{\psi}_{3i}
\\
g(t_{ij}|\psi_{0i},\psi_{1i},\psi_{2i},\nu) ~~~~~~ &\mbox{ if   } \tilde{\psi}_{3i} \leq t_{ij} \leq \tilde{\psi}_{3i} + \nu\\
\lambda_i + \psi_{2i}t_{ij} ~~~~~~  &\mbox{ if   } t_{ij} > \tilde{\psi}_{3i} + \nu 
\end{cases}
\end{split}
\end{equation}
\\
where $g$ is a third-degree polynomial function; $\tilde{\psi}_{3i}$ is the person-specific time when the smooth transition phase of length $\nu$ begins; and $\nu$ is an arbitrary value representing the time interval [$\tilde{\psi}_{3i}$; $\tilde{\psi}_{3i}$ + $\nu$] where the polynomial curve occurs, thus, if $\nu$ set to 0, it reduces to a PMM-abrupt model. $\psi_{0i}, \psi_{1i}$, and $\psi_{2i}$ have been previously defined for model \eqref{2}. 
\\
In order to be closer to the PMM-abrupt model defined in Equation \eqref{2}, the two linear parts in the PMM-polynomial model defined in Equation \eqref{3} should intersect at $\tilde{\psi}_{3i}+\frac{\nu}{2}$, corresponding to the middle of the transition phase and thus it is imposed that $\lambda_i=\psi_{0i}+\psi_{1i}(\tilde{\psi}_{3i}+\frac{\nu}{2})-\psi_{2i}(\tilde{\psi}_{3i}+\frac{\nu}{2})$. The continuity between the linear parts and the transition, as well as
 the smoothness of the transition function are ensured by considering the following constraints on the function $g$:

\begin{equation}\label{4}
\begin{split}
g(\tilde{\psi}_{3i}) &= \psi_{0i} + \psi_{1i}\tilde{\psi}_{3i}  
\\
g(\tilde{\psi}_{3i}+\nu) &= \lambda_i + \psi_{2i}(\tilde{\psi}_{3i}+\nu) 
\\
(\frac{\partial}{\partial t_{ij}}g)(\tilde{\psi}_{3i}) &= \psi_{1i}
\\
(\frac{\partial}{\partial t_{ij}}g)(\tilde{\psi}_{3i} + \nu) &= \psi_{2i}
\end{split}
\end{equation}

\noindent where $g$ is obtained by solving a system of four linear equations with four unknown parameters.  

\noindent As noted above, the person-specific intercept ($\psi_{0i}$) in Equations \eqref{2} and \eqref{3} represents the intersection of the first linear phase with the y-axis at death ($t=0$) and cannot be interpreted directly. Yet, the interpretation of the estimated mean and person-specific levels of cognition at death, that is the intersection of the second linear phase with the y-axis at time zero, is important information. As this requires a relatively simple reformulation, we re-wrote the PMM-polynomial defined in Equation \eqref{3} as follows: 

\begin{equation}\label{5}
\begin{split}
\mathbb{E}[y_{ij}] = 
\begin{cases}
\tilde{\psi}_{0i} + \psi_{1i}t_{ij} + (\psi_{2i}-\psi_{1i})(t_{ij}-\tilde{\psi}_{3i}+\frac{\nu}{2})
~~~~~~~ &\mbox{ if   } t_{ij} < \tilde{\psi}_{3i}
\\
g(t_{ij}|\tilde{\psi}_{0i},\psi_{1i},\psi_{2i},\nu) ~~~~~~ &\mbox{ if   } \tilde{\psi}_{3i} \leq t_{ij} \leq \tilde{\psi}_{3i} + \nu\\
\tilde{\lambda}_i + \psi_{1}t_{ij} ~~~~~~  &\mbox{ if   } t_{ij} > \tilde{\psi}_{3i} + \nu 
\end{cases}
\end{split}
\end{equation}
\\
and the constraints for a smooth transition defined in Equation \eqref{4} become

\begin{equation}\label{6}
\begin{split}
g(\tilde{\psi}_{3i}) &= \tilde{\lambda}_i + \psi_{2i}\tilde{\psi}_{3i}
\\
g(\tilde{\psi}_{3i}+\nu) &= \tilde{\psi}_{0i} + \psi_{1i}(\tilde{\psi}_{3i}+\nu) 
\\
(\frac{\partial}{\partial t_{ij}}g)(\tilde{\psi}_{3i}) &= \psi_{2i}
\\
(\frac{\partial}{\partial t_{ij}}g)(\tilde{\psi}_{3i} + \nu) &= \psi_{1i}
\end{split}
\end{equation}

\noindent where $\tilde{\psi}_{0i}$ is the person-specific level of global cognition at death, and the contraints $\tilde{\lambda}_i$ now corresponds to $\tilde{\psi}_{0i}+\psi_{1i}(\tilde{\psi_{3i}}+\frac{\nu}{2})-\psi_{2i} (\tilde{\psi}_{3i}+\frac{\nu}{2})$. The interpretation of $\psi_{1i}$, $\psi_{2i}$, $\tilde{\psi}_{3i}$, and $\nu$ in Equations \eqref{5} and \eqref{6} remains the same as in Equations \eqref{3} and \eqref{4}, respectively. In the following, we will only consider the re-formulated PMM-polynomial model \eqref{5}.

\newpage

\subsection{The sigmoidal mixed model with four parameters}
\begin{figure}[h]
\centerline{\includegraphics[width=270pt]{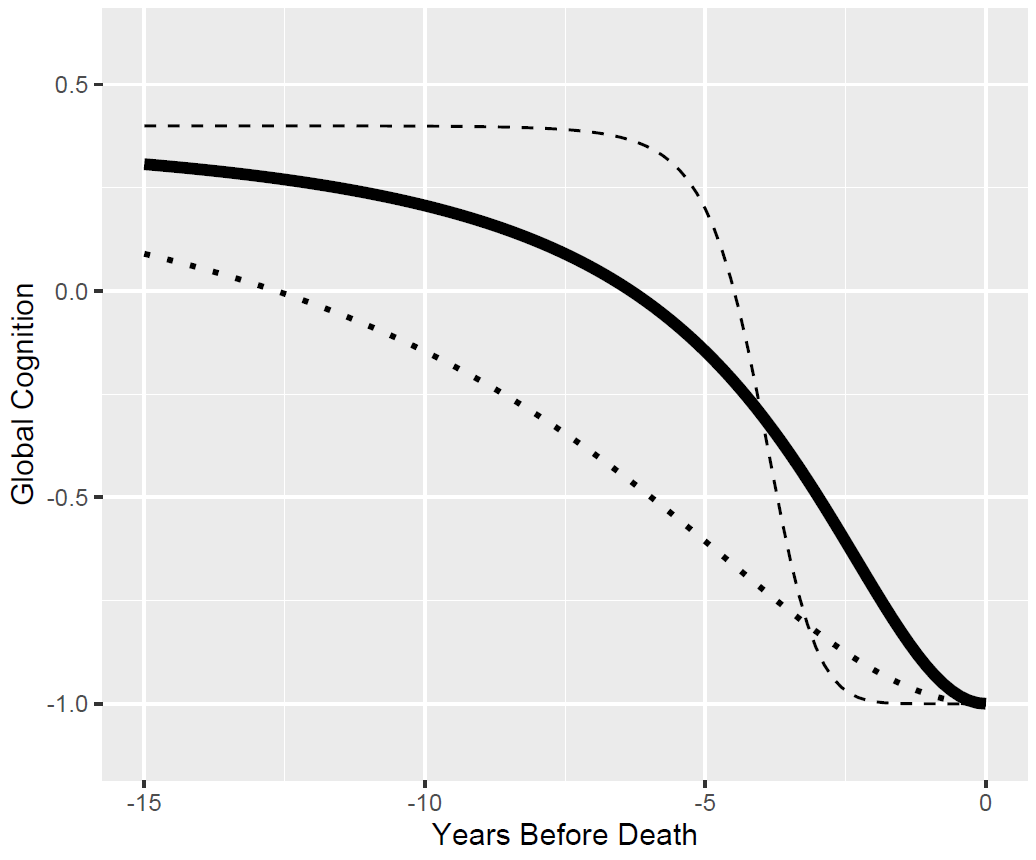}}
\caption{Example of possible marginal cogntiive trajectories based on sigmoidal mixed models for three distinct combinations of parameters ($\alpha_0$: final level at death; $\alpha_1$: initial level far from death; $\alpha_2$: half decline; $\alpha_3$: Hill slope). Solid line: $\alpha_0$=-1, $\alpha_1$=0.4, $\alpha_2$=4, $\alpha_3$=2. Dashed line: $\alpha_0$=-1, $\alpha_1$=0.4, $\alpha_2$=4, $\alpha_3$=8. Dotted line: $\alpha_0$=-1, $\alpha_1$=0.4, $\alpha_2$=8, $\alpha_3$=2. Adapated from Capuano and colleagues [20]. } \label{fig3}
\end{figure}

The SMM model [19] is a flexible non-linear mixed model, which allows for the analysis of non-linear trajectories with S-shaped (sigmoidal) forms, and can capture more linear trends. The SMM model described in this work is based on the four-parameter logistic model [20]. The SMM allows greater flexibility in fitting longitudinal outcomes. In particular, the SMM with four parameters allows for the inclusion of covariates related to four meaningful parametric quantities reflecting early levels, final levels, half decline, and variation of rate of decline (also known as the Hill slope) around the half decline time [20] (see Figure \ref{fig3}).
\\
Non-linear cognitive trajectories via this SMM can be formulated as follows: 

\begin{equation}\label{7}
\begin{split}
y_{ij} &= \psi_{1i} + (\psi_{0i} - \psi_{1i})  \underbrace{\frac{1}{\left(1 + (\frac{t_{ij}}{\psi_{2}})^{\psi_{3}}\right)}} + ~\epsilon_{ij} 
\\ &= \psi_{1i} + (\psi_{0i} - \psi_{1i}) \hspace*{1cm} p \hspace*{0.8cm} + \epsilon_{ij}
\end{split}
\end{equation}

\begin{center} 
with  
\end{center}

\begin{center}
~~~~~~ $\psi_{0i} = \alpha_0 + X_{0i}^\top\beta_0 + \eta_{0i}$, 
\vspace{3mm}
\\
~~~~~~ $\psi_{1i} = \alpha_1 + X_{1i}^\top\beta_1 + \eta_{1i}$,
\vspace{3mm}
\\
$\psi_{2} = \alpha_2 + X_{2i}^\top\beta_2$,
\vspace{3mm}
\\
$\psi_{3} = \alpha_3 + X_{3i}^\top\beta_3$,
\vspace{3mm}
\\
$\eta_i = (\eta_{0i}, \eta_{1i})^\top \sim \mathcal{N}(0, B)$ with $B$ = $\Bigl($ 
$\begin{matrix} 
\sigma_0 & \sigma_{01}
\\ 
\sigma_{01} & \sigma_1
\end{matrix}$
$\Bigl)$ ~ and ~ $\epsilon_{ij} \sim \mathcal{N}(0,\sigma_{\epsilon})$
\end{center}

\noindent This model \eqref{7} can be divided into two parts. The first part consists in the person-specific initial cognitive level far from death (i.e., $\psi_{1i}$). The second part involves the multiplication of the person-specific total cognitive change (i.e., ($\psi_{0i} - \psi_{1i}$); with $\psi_{0i}$ representing the person-specific final cognitive level at death) by a proportion of change (i.e., $p$), which is defined as a nonlinear function of time. In the second part, the fixed parameter $\psi_{2}$ denotes the point in time where half of the total decline occurred (i.e., half decline). In the context of cognitive decline before death, the domain of $\psi_{2}$ is restricted such that $\psi_{2}$<0. That way, when $t = \psi_{2}$, then $t/\psi_{2}=1$ and $p=0.5$. The fixed parameter $\psi_{3}$ corresponds to the Hill slope around the time $\psi_{2}$. In brief, the high value of $\psi_{3}$ decreases $p$ when measurement times precede the half decline (i.e., $t<\psi_2$), and increases $p$ when measurement times are beyond the half decline (i.e., $t>\psi_{2}$). $\alpha_0$, $\alpha_1$, $\alpha_2$, and $\alpha_3$ are the mean values for the final level at death, initial level, midpoint, and Hill slope, respectively; $X_{0i}$, $X_{1i}$, $X_{2i}$, and $X_{3i}$ are vectors of covariates associated with the vector of fixed effects $\beta_0$, $\beta_1$, $\beta_2$, and $\beta_3$, respectively; and $\eta_{0i}$ and $\eta_{1i}$ are person-specific random effects with $(\eta_{0i}, \eta_{1i})^\top \sim MVN(0, B)$ and $B$ assuming correlations between $\eta_{0i}$ and $\eta_{1i}$.

\subsection{Complementarity of the models}
The five mixed models assuming nonlinearity of change over time have different structures allowing to answer specific research questions. As cognitive decline is a progressive neurodegenerative process, which requires sufficiently long follow-up to be captured, the linearity assumption may be acceptable in the context of a short longitudinal observation period (e.g., $\leq$6 years) and/or when the number of repeated cognitive measures per subject is limited (e.g., $\leq$4 per individual). However, the accelerated decline may be observed over short periods, such as in prodromal dementia ($~$5 years before diagnosis)[6], and violation of this nonlinear assumption may lead to biased exposure-risk relationships [33]. As a result, the LMM-quadratic model is often used as a simple alternative to capture accelerations of cognitive change over time or to simply test its potential existence. Nevertheless, it is important to note that a quadratic time function can produce convex trajectories even though the underlying longitudinal process is likely to be monotonic, such as the decline of cognitive functions in older ages. Thus, another interesting alternative is the LMM-splines model, which, compared to LMM-quadratic, estimates the non-linear trajectory of cognitive paths directly from the data using a flexible basis of splines. The time function of LMM-splines is defined via a larger number of parameters compared to LMM-quadratic. The number of parameters depends on the number of nodes defined in the spline basis. The LMM-quadratic and LMM-splines are traditional mixed models estimating fixed effects of covariates, although in some contexts estimation their time-varying effects can be of interest. The FMM estimates the trajectory using splines and multiple time-varying coefficients.  Although the FMM is non-parsimonious, it represents a flexible way to visually inspect non-linear cognitive changes and effects over time. Despite their different structures, the LMM-quadratic, LMM-splines, and FMM models are constituted of a function of time defined through a linear combination of 2 or more parameters not directly interpretable. On the other hand, PMM and SMM models are two non-linear mixed models that offer a direct interpretation of four main parameters related to the evolution over time. The abrupt PMM model has a simpler interpretation and is useful but assumes two linear periods which is not necessarily realistic given the nature of cognitive decline with slow non-linear change over time. The smooth PMM helps to create a more believable trajectory. On the other hand, the SMM model is more complex but more realistic being smooth yet more flexible. The SMM model is more flexible and uniquely offers the possibility to assess parameters at the two extreme time points of the timescale of interest, as well as the decline midpoint and rate of decline. Overall, the convergence of these models is often conditioned by the amount of longitudinal information available over time, sample sizes, and the number of predictors specified. 

\section{ILLUSTRATIVE EXAMPLE}
\subsection{Data}
We applied and contrasted the interpretation of the models defined in Section 2 by investigating the association of education with cognitive decline before death for 1276 deceased participants from the Religious Orders Study (ROS) and Rush Memory and Aging Project (MAP) [34] with no dementia at study entry and who performed annual cognitive testing for at least 4 years before death (range=[4;26] years). To date, the follow-up rate exceeds 90\%. The main outcome considered was a composite score of global cognition calculated based on 17 psychometric tests, as described in detail elsewhere [34]. At baseline, the mean level of global cognition was -0.005 (SD=0.51). At the last assessment proximate to death, the mean cognitive level reached -0.872 (SD=1.13). The timescale considered was years before death, with an average follow-up of 11 years (SD=5) and a total number of observations of 13,064. The main predictors examined were sex (70\% of females) and self-reported years of formal education (mean=17 [SD=4] years). Other covariates studied were age at baseline (mean=80 [SD=7]) and age at death (mean=90 [SD=6]).

\subsection{Specification of the models}
Assuming an aligment at death, we modeled the cognitive trajectories using the following mixed models: 

\begin{equation}\label{14}
\begin{split}
\mbox{LMM-quadratic}: cog_{ij} = ~ &\alpha_{0} +  \beta_{01}edu_i +  \beta_{02}sex_i + \beta_{03}ageDeath_i + \eta_{0i} + \\
(&\alpha_1 + \beta_{11}edu_i + \beta_{12}sex_i + \beta_{13}ageDeath_i + \eta_{1i}) \times t_{ij} + \\
(&\alpha_2 + \beta_{21}edu_i + \beta_{22}sex_i + \beta_{23}ageDeath_i + \eta_{2i}) \times t^{2}_{ij} + \epsilon_{ij} 
\end{split}
\end{equation}

\begin{equation}\label{15}
\begin{split}
\mbox{LMM-splines}: cog_{ij} = ~ &\alpha_{0} + \beta_{01}edu_i + \beta_{02}sex_i + \beta_{03}ageDeath_i + \eta_{0i} + 
\\
&\left(\theta_{k1} + \theta_{k2}edu_i + \theta_{k3}sex_i + \theta_{k4}ageDeath_i + \eta_{ki}\right) \times \sum \limits_{k=0}^K B_k(t_{ij}) + \epsilon_{ij}
\end{split}
\end{equation}

\begin{equation}\label{18}
\begin{split}
\mbox{FMM}: cog_{ij} = \psi_{edu}(t_{ij})edu_i + \psi_{Sex}(t_{ij})sex_i + \psi_{ageDeath}(t_{ij})ageDeath_i + \eta_{0i}(t_{ij}) + \epsilon_{ij} 
\end{split}
\end{equation}

\begin{equation}\label{16}
\begin{split}
\mbox{PMM-polynomial}: \mathbb{E}[cog_{ij}] = 
\begin{cases}
\tilde{\psi}_{0i} + \psi_{1i}t_{ij} + (\psi_{2i}-\psi_{1i})(t_{ij}-\tilde{\psi}_{3i}+\frac{\nu}{2})
~~~~~~~ &\mbox{ if   } t_{ij} < \tilde{\psi}_{3i}
\\
g(t_{ij}|\tilde{\psi}_{0i},\psi_{1i},\psi_{2i},\nu) ~~~~~~ &\mbox{ if   } \tilde{\psi}_{3i} \leq t_{ij} \leq \tilde{\psi}_{3i} + \nu\\
\tilde{\lambda}_i + \psi_{1i}t_{ij} ~~~~~~  &\mbox{ if   } t_{ij} > \tilde{\psi}_{3i} + \nu 
\end{cases}
\end{split}
\end{equation}
\\
\noindent with

\begin{center}
$\tilde{\psi}_{0i} = \tilde{\alpha}_{0} + \tilde{\beta}_{01}edu_i + \tilde{\beta}_{02}sex_i + \tilde{\beta}_{03}ageDeath_i + \tilde{\eta}_{0i}$, 
\\\vspace{1.5mm}
$\psi_{1i} = \alpha_{1} + \beta_{11}edu_i + \beta_{12}sex_i + \beta_{13}ageDeath_i + \eta_{1i}$,
\\\vspace{1.5mm}
$\psi_{2i} = \alpha_{2} + \beta_{21}edu_i + \beta_{22}sex_i + \beta_{23}ageDeath_i + \eta_{2i}$, 
\\\vspace{1.5mm}
$\tilde{\psi}_{3i} = \tilde{\alpha}_{3} + \tilde{\beta}_{31}edu_i + \tilde{\beta}_{32}sex_i + \tilde{\beta}_{33}ageDeath_i + \tilde{\eta}_{3i}$, 
\\\vspace{1.5mm}
$\nu = 2$ ~~ and ~~ $\tilde{\lambda}_{i} = \tilde{\psi}_{0i}+\psi_{1i}(\tilde{\psi_{3i}}+\frac{\nu}{2})-\psi_{2i} (\tilde{\psi}_{3i}+\frac{\nu}{2})$
\\
\end{center}

\begin{equation}\label{17}
\begin{split}
\mbox{SMM}: cog_{ij} &=  
\psi_{1i} + (\psi_{0i} - \psi_{1i})  \frac{1}{\left(1 + (\frac{t_{ij}}{\psi_{2}})^{\psi_{3}}\right)} +  \epsilon_{ij} 
\end{split}
\end{equation}

\noindent with

\begin{center}
~~~~~~~~ $\psi_{0i} = \alpha_{0} + \beta_{01}edu_i + \beta_{02}sex_i + \beta_{03}ageDeath_i + \eta_{0i}$
\\
$\psi_{1i} = \alpha_{1} + \beta_{11}edu_i + \beta_{12}sex_i + \beta_{13}age0_i  + \eta_{1i}$
\\ 
$\psi_{2} = \alpha_{2} + \beta_{21}edu_i + \beta_{22}sex_i + \beta_{23}ageDeath_i$ 
\\ 
$\psi_{3} = \alpha_{3} + \beta_{31}edu_i + \beta_{32}sex_i + \beta_{33}ageDeath_i$
\end{center}

\noindent where $edu$ represents the years of education; $age0$ is the age at study entry, in years; $ageDeath$ is the age at death, in years; $sex$ is the sex. To facilitate the interpretation of the results, the continuous predictors $edu, age0,$ and $ageDeath$ were centered close to their means (values of 16, 80, and 90 years, respectively). For each model, $\psi_., \tilde{\psi_.}, \eta_., \tilde{\eta_.}, \theta_., \alpha_., \beta_., \nu, g, \lambda, \tilde{\lambda}$, and $\epsilon$ have been previously defined in Section 2. $B_k(t_{ij})$ in the LMM-splines model \eqref{15} is a basis of natural cubic splines with 3 inner knots at the $25^{th}$, $50^{th}$, and $75^{th}$ percentiles of the time before death (the number of nodes was determined by BIC and gain in flexibility between marginal estimated trajectories; Supplementary Figure 1A). For the FMM model \eqref{18}, we applied cubic B-splines with 9 equal-quantile inner knots to generate the transformed spline basis functions. For the SMM model  \eqref{17} as it uniquely includes a parameter related to the early level of global cognition ($\psi_1$),  we naturally adjusted this parameter on the age at study entry ($age0$), while all other parameters were adjusted for $ageDeath$.

\subsection{Fitting the models}
All the models defined in Section 3.2. were fitted using the $\textbf{R}$ software version 4.0.3. For LMM-quadratic and LMM-splines we used the $\textbf{hlme}$ function of $\textbf{lcmm}$ package [35] version 1.7.8., which applies the maximum likelihood framework. In particular, $\textbf{lcmm}$ allow us to fit LMM-splines without difficulties when using the package $\textbf{splines2}$ [36] in combination -- unlike with $\textbf{lme4}$ [37], for example. For FMM, the parameters and penalty terms were estimated simultaneously by REML method, using the $\textbf{lme}$ function from the $\textbf{nlme}$ package [38]. Lastly, for PMM-polynomial and SMM, we fitted the models using $\textbf{nlive}$ package [39] version 0.1.0; the computational technique for maximum likelihood estimation implemented in $\textbf{nlive}$ is the Stochastic Approximation Expectation Maximization algorithm [40].
\\
In general, all models converged without difficulty over a reasonable processing time (<173 seconds), using a computer containing an i7-8565U processor and 16 gigabytes of RAM. Because the specification and the estimation procedures of the models can differ, formal comparisons cannot be performed. Nevertheless, to examine the goodness of fit of the models, we examined the estimated parameters and relied on the BIC, with lower BIC indicating a relatively better fit. In addition, we plotted and compared cognitive trajectories estimated before death according to extreme educational level (low [14 years] versus high [19 years]), in the analytical sample and in randomly selected participants. All data in these analyses (and descriptions of the studies and variables) can be requested through the Rush Alzheimer’s Disease Center Research Resource Sharing Hub at www.radc.rush.edu. The R code to replicate the analyses of this study can be provided by the corresponding author upon reasonable request.

\subsection{Results}
\subsubsection{Estimated parameters}
Table 1 displays the fixed effects estimated for LMM-quadratic, LMM-splines, PMM-polynomial, and SMM for the most common profile of covariates in the analytical sample. For FMM, we represented in Figure \ref{fig3} the mean curves estimated and 95\% confidence intervals in the 15 years before death. For LMM-quadratic both linear and quadratic time terms were negative and significant (all $P$$<$0001; Table 1), indicating that cognitive decline accelerated before reaching a mean level of -1.17 units at death. Each additional year of education was related to 0.03 units increase in cognitive function at death ($P$=.006; Table 1). Lastly, although each interaction term between education, time, and time squared was nonsignificant (all $P\geq$.1; Table 1), we found that education was related to the overall quadratic trajectory of global cognition when we performed a multivariate Wald test for combinations of parameters with the null hypothesis $H_0$: $edu \times t = edu \times t^{2} = 0$ ($P$=.01).

\noindent For LMM-splines, we selected the model with 3 inner nodes at quantiles (Supplementary Figure 1A). In Table 1, the estimated parameters of the selected model $\theta_{.1}, \theta_{.2}, \theta_{.3}$, and $\theta_{.4}$ (subscript . refers to 1 for the intercept and refers to 2, 3, 4 for those related to interactions with $education$, $sex$, $ageDeath$, respectively) are related to the basis functions of natural splines $B_1(t)$, $B_2(t)$, $B_3(t)$, and $B_4(t)$, respectively. By representing the natural splines basis used in the model, we can see that $B$ can be different from zero at the time of death: $B_1(0)$ = 0, $B_2(0)$ = -0.46, $B_3(0)$ = 0.51, and $B_4(0)$ = 0.95 (Supplementary Figure 1B); thus, precluding direct interpretation of the mean level at death, or the intercept. Alternatively, the mean cognitive level at death can be calculated a posteriori as follows: $\hat{\alpha}_0$ + ($\sum_{k=1}^4 + \hat{\theta}_{1k}B_k(t=0)$) = 0.29 -- (0.40$\times$0) + (0.64$\times$0.46) -- (1.16$\times$0.51) -- (1.37$\times$0.95) = --1.31 units. Lastly, education was not related to the overall cognitive trajectory when using a multivariate Wald-test for combinations of parameters with the null hypothesis $H_0$: $edu \times \theta_{21} = edu \times \theta_{22} = edu \times \theta_{23} = edu \times \theta_{24}$ = 0 ($P$=.2).

\noindent For PMM-polynomial, the mean estimated level of global cognition at death was -1.27 standard unit (Table 1). Each additional year of education was related to an increased final cognitive level of 0.03 units ($P$=.01; Table 1). From study entry to death, global cognition declined an average of 0.04 units per year until an average of 4.25 years before death when the mean rate of decline accelerated to 0.35 units per year, a 10-fold increase. Higher education was related to earlier changepoint. Ther was no association between education and the preterminal and terminal decline (all $P$$>$.1; Table 1). 

\noindent For SMM, the initial and final levels of cognition were estimated to be 0.60 and -1.38 units, respectively (Table 1). Half of the decline in global cognition was estimated to occur 3.95 years before death. In other words, half of the total loss in cognition will occur on average in the last ~4 years of life. Further, each additional year of formal education was related to higher initial (estimate=0.04, $P$<.0001; Table 1) and final (estimate=0.04, $P$=.002; Table 1) scores of cognition, as well as  to the decline midpoint (estimate=0.05, $P$=.001; Table 1) and the rate of decline (estimate=0.02, $P$=.003; Table 1). 

\begin{figure}[h]
\centerline{\includegraphics[width=390pt]{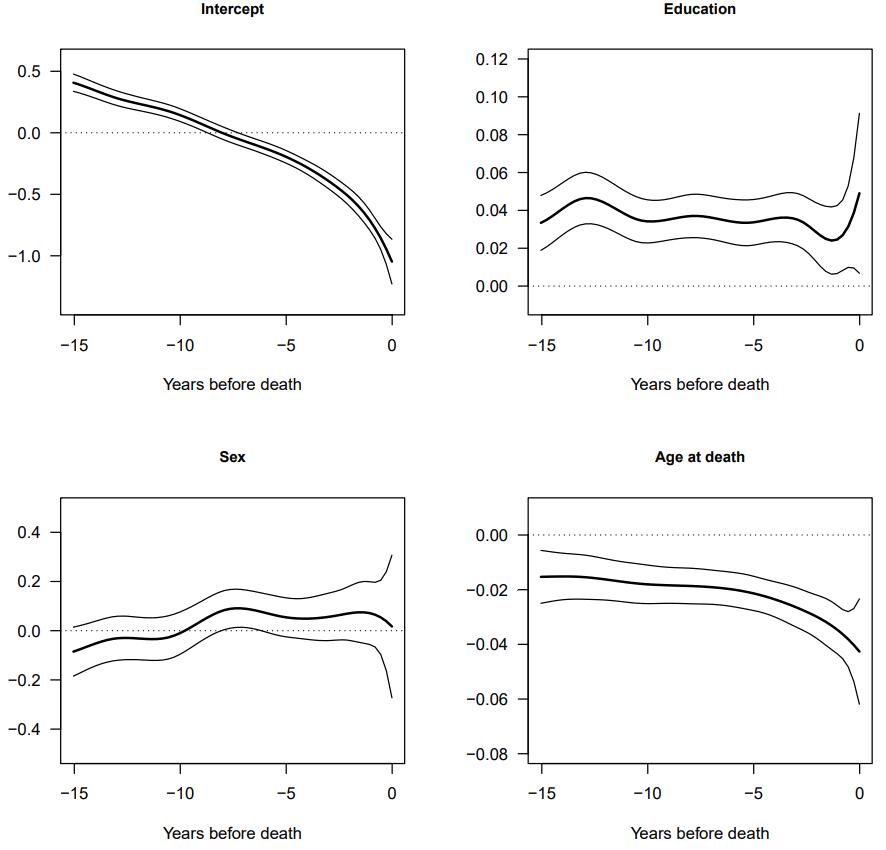}}
\caption{Estimated fixed functions and 95\% point-wise confidence intervals obtained under the functional mixed model with time-varying coefficients in the 15 years before death, ROSMAP (n=1276). Education and age at death were centered at 17 and 90 years, respectively. Dotted lines are the horizontal at zero. }\label{fig4}
\end{figure}

\noindent Finally, for FMM, the intercept function represents the marginal cognitive trajectory in the 15 years before death, in the reference group (Figure \ref{fig4}). The mean cognitive level was about 0.4 units and decreased gradually until death to reach about -1.1 units. For education, the time-varying coefficient estimates were significant and remained relatively constant over time, with each additional year of education associated with approximately 0.04 unit higher cognitive levels. 

\noindent Although formal comparisons between models cannot be performed, when evaluating the statistical fit of the models based on the BIC, the best goodness of fit was achieved first by the LMM-splines model (9510.33), followed by the FMM (10004.41),  the SMM (10821.82), the LMM-quadratic (10933.46) and the PMM-polynomial (12440.1).

\newpage

\begin{landscape}
\noindent Table 1. Estimated fixed parameters from each mixed model, excluding the functional mixed model, ROSMAP (n=1276). Education and age at death were centered at 17 and 90 years, respectively. 
\\
\tiny
\begin{tabular}{lcclcclcclcc}
\\
\toprule

Model term & 
\multicolumn{2}{c}{LMM-quadratic} & 
Model term & 
\multicolumn{2}{c}{LMM-splines} & 
Model term & 
\multicolumn{2}{c}{PMM-polynomial} & 
Model term & 
\multicolumn{2}{c}{SMM} 

\\
\cline{2-3}
\cline{5-6}
\cline{8-9}
\cline{11-12}

& 
Estimate (SE) & \textit{P} & & 
Estimate (SE) & \textit{P} & & 
Estimate (SE) & \textit{P} & & 
Estimate (SE) & \textit{P}

\\
\midrule

\\
Intercept & -1.17 (0.05) & $<$.0001 & 
Intercept &  0.29 (0.02) & $<$.0001 &   
Last leve (intercept) &              &        &   
Last level (intercept) &              &        

\\

\textit{edu} & 0.03 (0.01) & .006 &   
\textit{edu} &  0.05 (0.004) & $<$.0001 &  
~~~~Mean &  -1.27 (0.05) & $<$.0001 &    
~~~~Mean &  -1.38 (0.06) & $<$.0001 

\\

\textit{sex} & 0.31 (0.09) & $<$.0003 &   
\textit{sex} &  -0.14 (0.03) & $<$.0001 &  
~~~~\textit{edu} &  0.03 (0.01)  & .01   & 
~~~~\textit{edu} &  0.04 (0.01)  & .002  

\\
\textit{ageDeath} & -0.04 (0.01) & $<$.0001 &  
\textit{ageDeath}       & -0.01 (0.003) & .004   &   
~~~~\textit{sex}              & 0.30 (0.08)  & .0001 &   
~~~~\textit{sex}             & 0.28 (0.11)  & .01

\\
$t$    & -0.22 (0.01) & $<$.0001 &     
$\theta_{11}$ & -0.40 (0.03)  & $<$.0001 &   
~~~~\textit{ageDeath}              & -0.04 (0.01)  & $<$.0001 &     
~~~~\textit{ageDeath}             & -0.04 (0.01)  & $<$.0001 

\\
$t^2$   & -0.01 (0.001) & $<$.0001 &     
$\theta_{12}$ & -0.64 (0.03) & $<$.0001 &
 Slope 1     &              &        &    
Midpoint                 &              &

\\
\textit{edu} $\times$ $t$ & 0.0003 (0.002) & .9 &     
$\theta_{13}$ & -1.16 (0.05)  & $<$.0001 &   
~~~~Mean                 & -0.04 (0.003) & $<$.0001 &
~~~~Mean            &  3.95 (0.08)  & $<$.0001

\\
\textit{edu} $\times$ $t^2$ & 0.0001 (0.0001) & .2 &     
$\theta_{14}$ & -1.37 (0.04)   & $<$.0001 &   
~~~~\textit{edu}            & 0.001 (0.001) & .1     & 
~~~~\textit{edu}            & 0.05 (0.02)    & .001

\\
\textit{sex} $\times$ $t$ & 0.05 (0.01) & .001 & 
\textit{edu} $\times$ $\theta_{21}$ & -0.01 (0.01)   & .02 &  
   ~~~~\textit{sex}                     & 0.02 (0.01) & .001 &  
~~~~\textit{sex}                     & -0.85 (0.18) & $<$.0001

\\
\textit{sex} $\times$ $t^2$  & 0.001 (0.001) & .1 &    
\textit{edu} $\times$ $\theta_{22}$ & -0.01 (0.01)   & .1     & 
~~~~\textit{ageDeath}                     & -0.0002 (0.0004) & .1 &  
~~~~\textit{ageDeath}                    &  0.03 (0.01)   &  .01

\\
\textit{ageDeath} $\times$ $t$   & -0.003 (0.001) & .005 &         
\textit{edu} $\times$ $\theta_{23}$ &  -0.02 (0.01) & .1     &  
Changepoint                      &                &        &
Hill slope                       &               &

\\
\textit{ageDeath} $\times$ $t^2$  & -0.0001 (0.0001) & .4 &   
\textit{edu} $\times$ $\theta_{24}$ & -0.02 (0.01) & .1     &
 ~~~~Mean                         &  -4.25 (0.10) & $<$.0001 &  
~~~~Mean                    & 1.27 (0.03)  & $<$.0001

\\
  &  &  &     
\textit{sex} $\times$ $\theta_{31}$ & 0.20 (0.05) & $<$.0001   &
~~~~\textit{edu}                   & -0.08 (0.02) & .001     &  
~~~~\textit{edu}                   &  0.02 (0.01)  & .003

\\
  &  &  &     
\textit{sex} $\times$ $\theta_{32}$ & 0.27 (0.06) & $<$.0001 &
~~~~\textit{sex}                    & 0.47 (0.19) & .01     &  
~~~~\textit{sex}                    & -0.24 (0.06)  & .1

\\
  &  &  &     
\textit{sex} $\times$ $\theta_{33}$ & 0.42 (0.09) & $<$.0001 & 
~~~~\textit{ageDeath}                    & -0.11 (0.01)  & $<$.0001 & 
~~~~\textit{ageDeath}                    & 0.004 (0.004)  & .1

\\
 &  &  &     
\textit{sex} $\times$ $\theta_{34}$ & 0.40 (0.08)   & $<$.0001 &
 Slope 2                      &                &        &  
Initial level                       &               &

\\
 &  &  &     
\textit{ageDeath} $\times$ $\theta_{41}$ & -0.01 (0.004) & .03   & 
~~~~Mean & -0.35 (0.01) & $<$.0001 &   
~~~~Mean & 0.60 (0.03) & $<$.0001

\\
 &  &  &     
\textit{ageDeath} $\times$ $\theta_{42}$ & -0.02 (0.004) & .0004   &  
 ~~~~\textit{edu}                   & -0.002 (0.003) & .3     &  
 ~~~~\textit{edu} & 0.04 (0.01) & $<$.0001

\\
 &  &  &      
\textit{ageDeath} $\times$ $\theta_{43}$ & -0.03 (0.01) & $<$.0001 &  
~~~~\textit{sex}  & 0.05 (0.02) & .03  &  
~~~~\textit{sex} & -0.17 (0.05) & .0002

\\
 &  &  &      
\textit{ageDeath} $\times$ $\theta_{44}$ & -0.03 (0.01) & $<$.0001 &  
~~~~\textit{ageDeath}  & 0.0001 (0.002) & .5  & 
~~~~\textit{age0} & -0.01 (0.003) & $<$.0001

\\
\bottomrule

\end{tabular}
\end{landscape}

\newpage

\subsubsection{Estimated cognitive trajectories according to educational level}
\begin{figure}[h]
\centerline{\includegraphics[width=450pt]{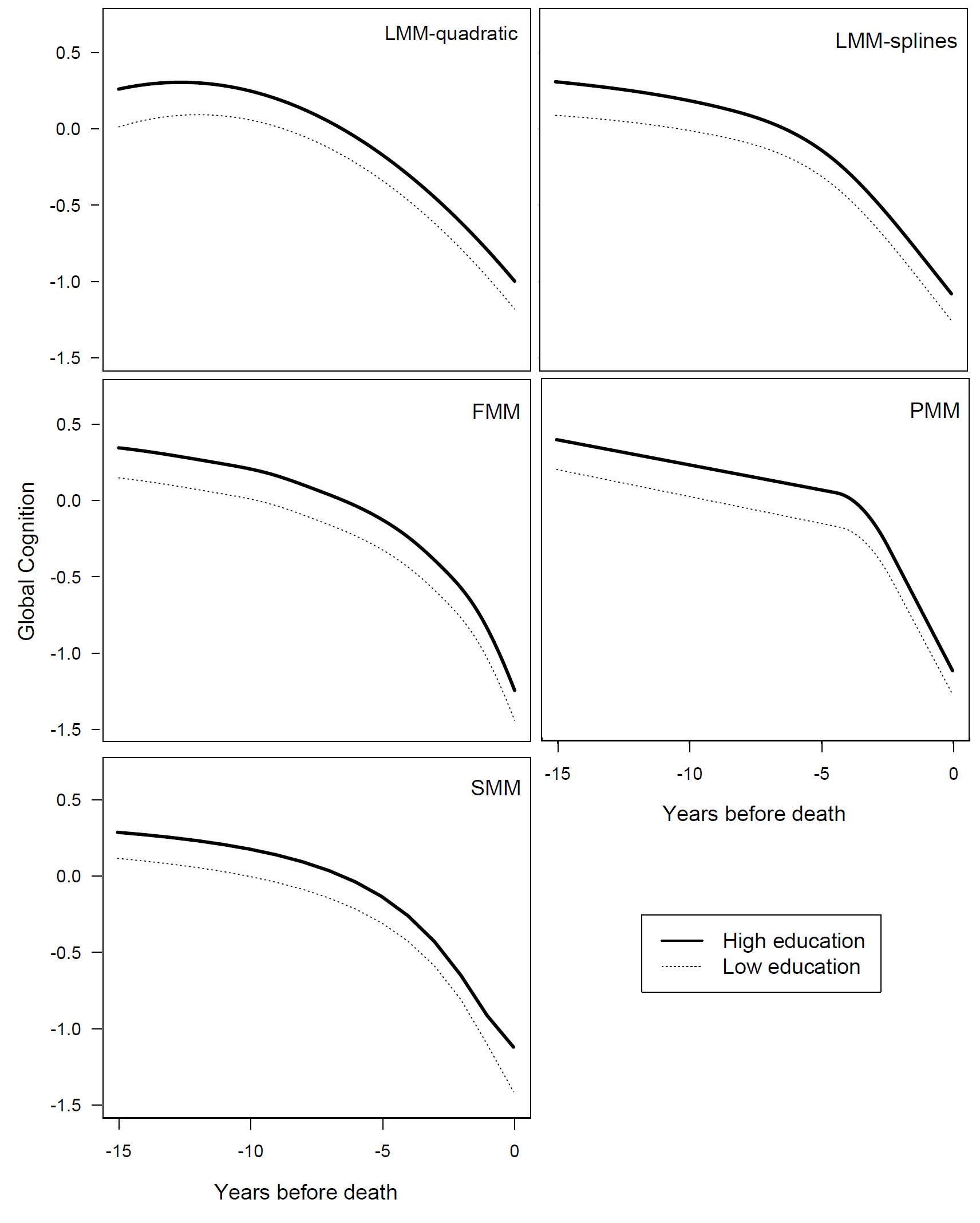}}
\caption{Mean trajectories of global cognition estimated before death under the mixed models in typical participants (i.e., woman aged 90 at time of death) with low (25$^{th}$ percentile, 14 years) versus high (75$^{th}$ percentile, 19 years) levels of education, ROSMAP (n=1276). }\label{fig5}
\end{figure}

To help interpretation of the parameters related to time, it can be helpful to represent graphically the mean estimated trajectories over time. In particular, for example, it can help to identify the time when occurring an acceleration or a crossing of trajectories. In this work, with an alignment at death, we compared the marginal estimated trajectories of global cognition in the years before death among typical participants with low versus high levels of education (Figure \ref{fig5}). 

\begin{figure}[h]
\centerline{\includegraphics[width=420pt]{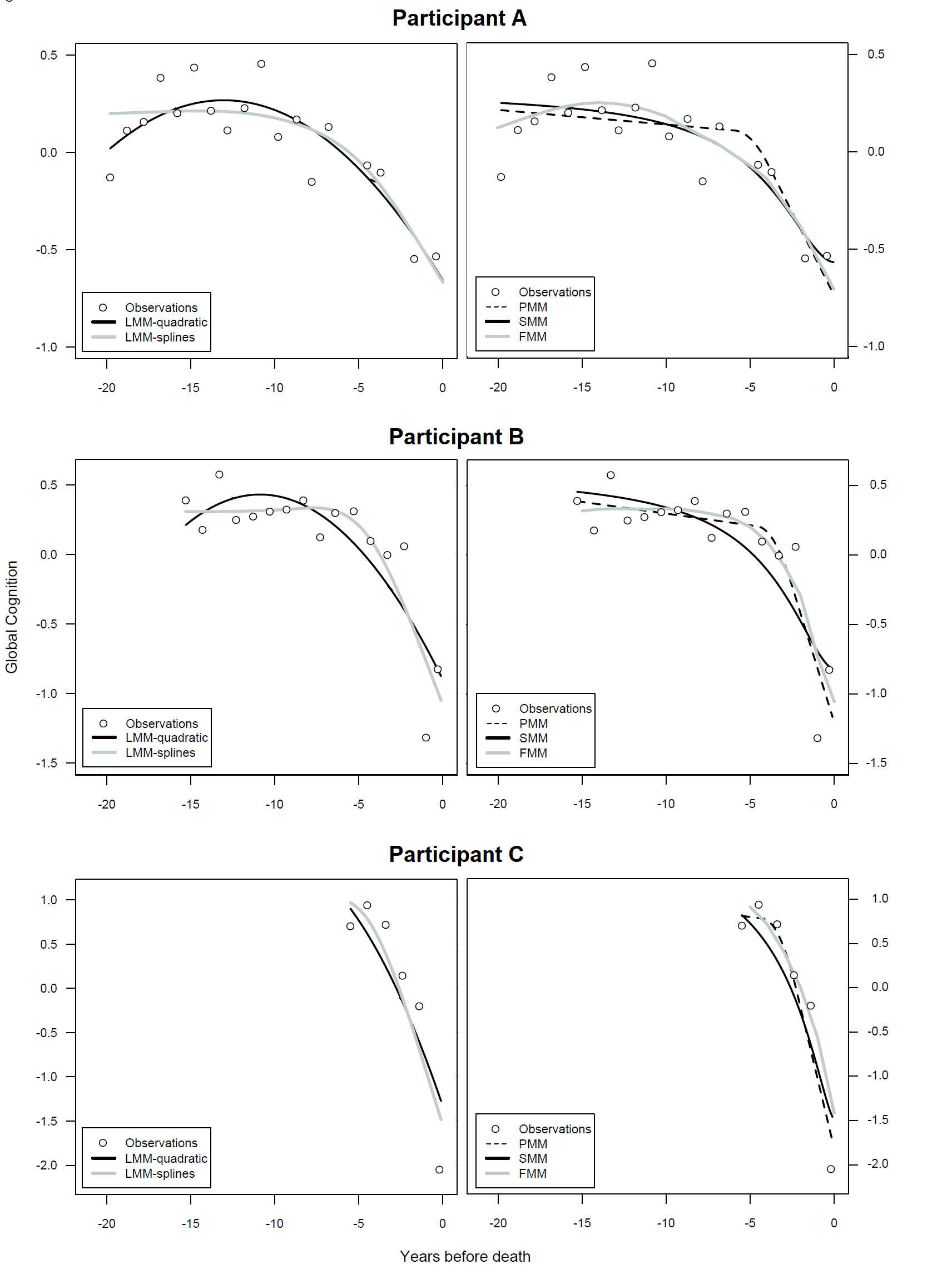}}
\caption{Person-specific trajectories of global cognition estimated before death under the mixed models for three participants with approximately 20, 15 , and 5 years of follow-up.} \label{fig6}
\end{figure}

\noindent Trajectories obtained for PMM are as expected based on the previous interpretation of the parameters; individuals with higher education had better initial cognitive levels but similar cognitive decline over time than individuals with low educational levels. In contrast, it is interesting to see that for LMM-quadratic, trajectories of participants with low and high educational attainment were characterized by a slightly unrealistic increase from -15 to -13 years before death, followed by an accelerated cognitive decline until death. In addition, for LMM-splines, PMM-polynomial, and SMM, estimated cognitive trajectories were characterized by a steady decrease over time followed by a point of inflection observed approximately 5 years before death, reflecting the onset of accelerated terminal decline. Lastly, the FMM displayed a relatively early accelerated decline (Figure \ref{fig5}). 
\\
In addition, the representation of the person-specifc cognitive trajectories can provide valuable insights on the model adjustments to the individual data. In Figure \ref{fig6}, we plotted the repeated observations of global cognition, as well as the cognitive trajectories estimated from all the mixed models for three randomly selected participants with different lengths of follow-up (i.e., 20, 15, and 5 years before death). For LMM-quadratic, a non-realistic U-shaped curve is observed for participants with moderate and long follow-up duration. Otherwise, the estimated trajectories reflect relatively good fits over time.

\newpage

\newpage

\newpage

\section{Simulation study}
We conducted a simulation study to formally assess the empirical properties of the five mixed models capturing non-linearity of change over time. First, we examined their empirical properties assuming 5\% missing data and two underlying types of trajectory. We assessed the estimation accuracy of the marginal values before death, year by year, using the empirical mean squared error $MSE(t)=\frac{1}{R} \sum_{r=1}^{R}\left(\hat{Y}_r(t) - Y(t)\right)^{2}$ and the bias $bias(t)=\frac{1}{R} \sum_{r=1}^{R}\hat{Y}_r(t) - Y(t)$, where $\hat{Y}_r(t)$ and  $Y(t)$ represent the estimated level of cognition and the underlying true level, respectively, at year $t$ before death for $r$ replicates ($r$=1,...,500). Second, we challenged the models imposing different patterns of missingness, assessment intervals and sample sizes. 

\subsection{Data generation}
We simulated data close to the longitudinal cognitive trajectories observed in the ROSMAP [34], using the flexible sigmoidal structure with four parameters (see Section 2.5.). Decedent-specific visit times were generated using a uniform distribution in [–2, 2] months around theoretical annual visits from –24 years to death (year 0), and considering an average follow-up of 10 years (SD=5). The 5\% proportion of missing visits completely at random was selected to mimic the ROSMAP. 
\\
Two main simulation scenarios were considered. In Scenario A, the marginal trajectory of global cognition before death was characterized by an early, progressive cognitive decline over time and a moderate terminal decline ($\beta_0$=–1.03, $\beta_1$=0.37, $\beta_2$=–4.0, $\beta_3$=1.69). In Scenario B, the marginal trajectory of global cognition was characterized by a late and more pronounced accelerated terminal decline ($\beta_0$=–1.03, $\beta_1$=0.37, $\beta_2$=–2.50, $\beta_3$=2.50). In both Scenarios A and B, we assumed a variance of 2.13 ($\sigma$=1.46) and 0.26 ($\sigma$=0.51) for the random effects $b_0$ and $b_1$, respectively, a correlation of zero between random effects, and an error variance of 0.08 ($\sigma$=0.28). As an illustration, the Supplementary Figure 2 displays boxplots of the cognitive values generated before death in Scenarios A and B.
\\
Using data generated in Scenarios A and B, we considered four main challenges commonly encountered in longitudinal studies of cognitive aging:
\begin{itemize}
    \item Scenarios A1 to B1: larger proportion of missing cognitive data in the year before death (30\%), mimicking ROSMAP
    \item Scenarios A2 to B2: half of individuals (n=500) had only 4 cognitive measures prior to death
    \item Scenarios A3 to B3: cognitive measures were assessed every 3 years instead of every year
    \item Scenarios A4 to B4: smaller study sample size (n=150)
\end{itemize}

\noindent For each Scenario, we simulated 500 datasets of 1,000 individuals each (except Scenarios A4-B4 with a smaller number of  individuals). For each generated dataset, we fitted the six mixed models and represented graphically the true and the average of the 500 estimated marginal cognitive trajectories in the 15 years before death.

\subsection{Empirical properties}

\begin{figure}[h]
\centerline{\includegraphics[width=280pt]{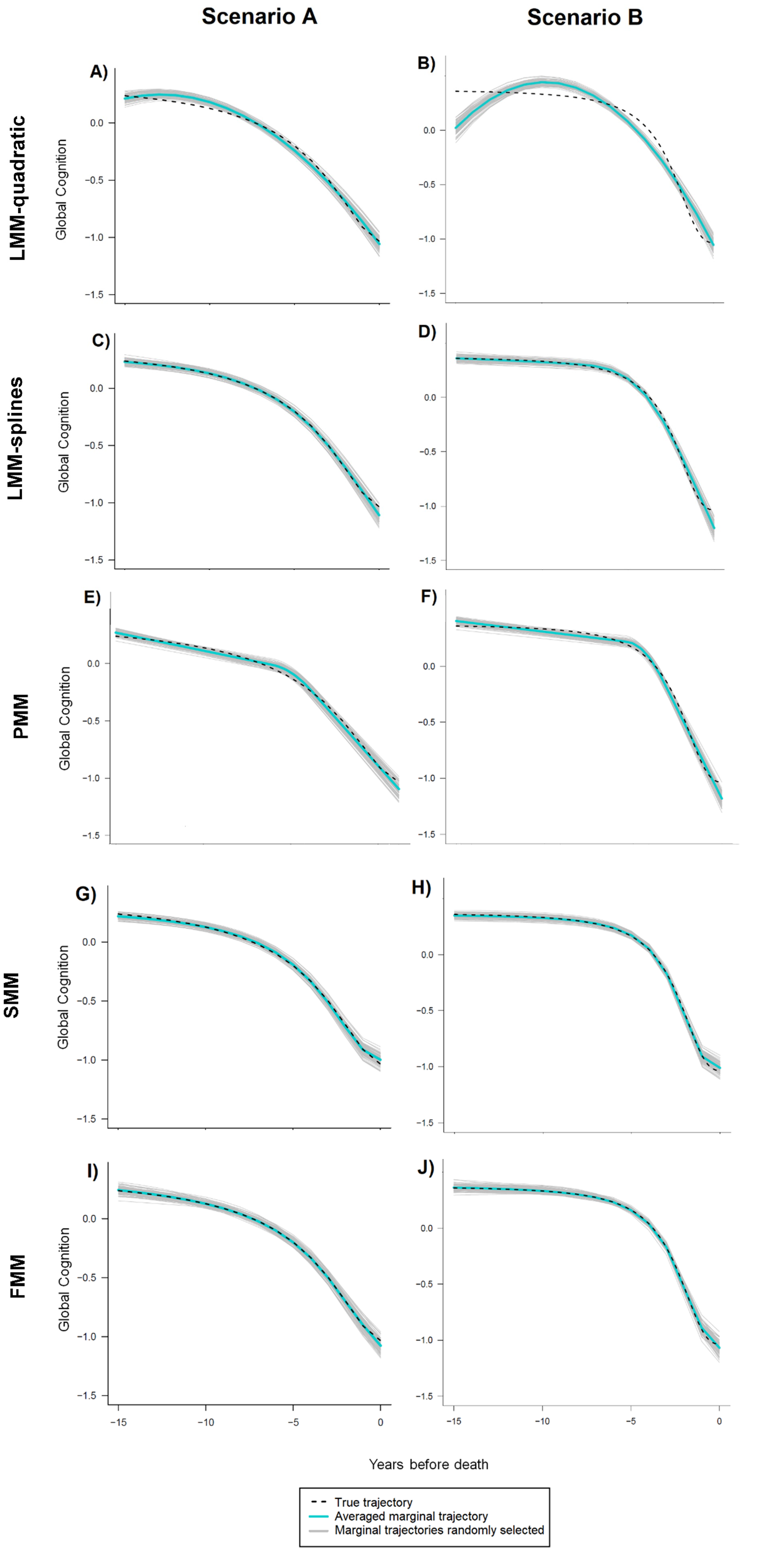}}
\caption{Average of the marginal trajectories of global cognition estimated before death across 500 simulations of 1,000 individuals each (blue line) and 100 marginal trajectories randomly selected (grey lines) for each mixed model in Scenarios A and B.}\label{fig7}
\end{figure}

Overall, we did not experience any difficulties in fitting the models. Only convergence difficulties emerged in Scenario B3 for the SMM model, that was resolved by changing the starting values. In Scenarios A and B, the LMM-splines, PMM-polynomial, SMM, and FMM models generally provided estimates with low bias (Figure \ref{fig7}C to J); we only found slight deviations from the true underlying trajectory of global cognition in the 2 years before death (all MSE$\leq$0.031 and Bias in the range of -0.167, 0.104; Supplementary Table 1). In addition, for the PMM-polynomial model, we generally observed slight deviations when the smooth transition occurred (Figure \ref{fig7}E–F); as expected, the higher the rate of terminal decline (Scenario B), the shorter the duration of the transition and deviations (Figure \ref{fig7}E-F). Finally, for the LMM-quadratic model, although the bias was overall lower than expected (all MSE$\leq$0.116 and Bias in the range of -0.337,0.140; Supplementary Table 1), the trajectories upstream of death showed an unrealistic increase in cognitive ability (Figure \ref{fig7}A–B). Notably, for Scenario B, we observed a bell-shaped curve intersecting the true marginal trajectory at -13 and -6 years before death (Figure \ref{fig7}D). Biases for Scenarios A and B were further challenging the models by (a) excluding 30\% of cognitive values collected in the year before death (b) decreasing the study sample size to 150 individuals, (c) increasing the spacing between follow-up visits to every 3 years, (d) limiting half of the individuals to having only 4 cognitive measures before death. Challenges (a) to (d) are illustrated in Supplementary Figures 3 to 6, respectively. Bias and MSE are provided in Supplementary Tables 2 to 5, respectively. The LMM-quadratic generally presented high bias. For the other models, challenges (a), (b) and (d) provided small bias but the challenge (c) increased the bias in the last year close to death that was more marked for SMM and moderate for LMM-splines, PMM-polynomial, and FMM. 

\subsection{Recommendations}
Simulation and real data examples contrasted the five alternative mixed models, illustrating the superiority of fit and richness of interpretation of non-linear modeling. The LMM-quadratic clearly underperformed compared with the other models, especially when the duration of follow-up was 10 years or more. Notably, LMM-quadratic provided unrealistic cognitive trajectories in late life, with improvement followed by decline in global cognition before death, and biases were accentuated in the case of marked terminal decline. Therefore, we recommend the use of this quick and easy-to-fit model only for exploratory purposes, in order to detect possible accelerated nonlinear changes over time. If any, the use of more sophisticated mixed models, such as the LMM-splines, should be applied when sufficient longitudinal data are available to allow the fitting of these models. 
\\
The LMM-splines, PMM-polynomial, SMM, and FMM, performed well in the different Scenarios of interest. In this work we tested the performance of the models with a smaller sample of 150. Smaller samples did not impose a problem as long as further challenges were not present. That tells us that, in practice, this is a minimum sample that should be used. Similar with less frequent follow-up. Models perform well with annual data challenged with different patterns of missingness. However further spacing in follow-up (here the highest spacing testing was 3 years) increased bias even when no further challenges were imposed.
\\
Overall, and in line with prior research, LMM-splines, FMM, PMM-polynomial, and SMM are our recommended approaches for comprehensively characterizing how specific covariates influence the longitudinal outcome time course. In particular, PMM and SMM may be viewed as parsimonious approaches that allows for the interpretation of specific quantities. Thus, although the ideal conditions ensure that accuracy of non-linear mixed models may be rare and difficult to ascertain, they should be conducted whenever feasible.

\newpage

\section{Discussion}
In this study, we examined five mixed models to characterize late-life cognitive trajectories from baseline to death. The models considered in this work have different structures. By considering a quadratic term for time (LMM-quadratic) or by approximating the function of time with splines (LMM-splines), the LMM allows for a non-linear change in time. The piecewise linear mixed model (PMM) with a random changepoint separates the trajectory into two linear phases and allows for the estimation of the onset of terminal decline (the point in time between phases). The sigmoidal model (SMM) with four parameters enables estimation of early and last cognitive level, decline midpoint, and rate of decline. Lastly, the FMM is a spline model with multiple time-varying coefficients that, although non-parsimonious, is a flexible way to visually inspect non-linear cognitive change over time. Overall, the application and simulations aligned with prior research work and reinforced that assuming linearity was not reasonable with these data and this approach should be reserved for special situations where simplicity is needed over precision. Among the non-linear mixed models, LMM-quadratic was the one with the worse fit when mimiking the ROSMAP data and is not recommended in other cohorts with relatively similar study design. Otherwise, models had a good performance. 
\\
Overall, the aforementioned approaches for longitudinal data have the potential to improve our ability to characterize cognitive aging and identify risk factors for cognitive decline without heavy computational demand. Models were fitted in R language for Statistical Computing [41] and we considered R packages with relatively user-friendly functions and a variety of options, although other packages exist for LMMs [37] and PMMs with abrupt [30,42] or Bacon Watts transition [30]. The codes to fit the PMM-polynomial with polynomials and the SMM were newly developed in this work and were found to perform well in the simulation. In general, all the models were fitted without difficulty. In addition, the convergence time was relatively reasonable, although we noticed that LMM-splines, PMM, and FMM required more time to converge (range: 90 to 173 seconds) compared with SMM (<20 seconds). In other settings, convergence time can increase with larger study sample sizes and/or a higher number of predictors considered. For LMM-splines and FMM, an increasing number of internal nodes can substantially impact the convergence time. 
\\
The choice of the model can be defined according to specific aims and study settings. In the first examinations of the non-linear cognitive trajectories, LMM-splines can be an easy model to fit with high flexibility. The FMM also provides high flexibility and can be used to graphically inspect if the effect of the covariate is uniform over time. In contrast, the PMM with two linear phases is widely used to estimate directly and readily the onset of terminal decline, as well as the pre-and post-decline. Alternatively, SMM can be used when there is no desire to impose a changepoint and when risk factors for both first and last cognitive levels want to be studied, along with factors that influence an earlier half-decline. When examining different settings in simulations, we observed that larger sample sizes (n>=150) were required for SMM and PMM to facilitate the convergence of the models compared with LMM-splines and FMM. Overall, assuming a very basic model with no predictors, greater sample sizes than 150 could tolerate more spacing of follow-up and more missingness close to death. 
\\
The models were applied to the ROSMAP data to study the association between education and global cognition before death. Education is an important marker for cognitive reserve [43]; high cognitive reserve in highly educated individuals could help delay the onset of terminal decline. Our findings suggest that higher education was associated with higher cognition at baseline in the SMM model and death in all the models [not always observed for both]. However, education was generally unrelated to the annual rate of cognitive decline before death, suggesting that the contribution of education to the cognitive reserve hypothesis [44] is limited to its association with premorbid cognitive change. In addition, the graphics of FMM suggested that education has a uniform effect on cognition over time. The lack of association of education with terminal decline is consistent with previous longitudinal studies using either standard LMM with a linear function of time [45], PMM [9,46,47], or SMM models [20]. In two prior ROSMAP studies, education was associated with an earlier changepoint [47] and a slower rate of decline [20], while results were conducted on smaller sample sizes and shorter observation periods.
\\
Our study presents limitations in scope. The mixed models considered in this work assume that the longitudinal outcome of interest is continuous and approximately Gaussian. While the illustrative example meets this assumption, applying these models to non-normal data can lead to misleading results [48]. For example, the 30-point MMSE score [49], which is widely used in population-based longitudinal studies to quantify global cognitive function, has poor metrological properties, including notable floor/ceiling effects and curvilinearity [50]. While not covered here, alternative approaches can be used, including pre-transformations of the non-Gaussian outcome [51]  or the use of the latent process mixed model for non-Gaussian outcome [52]. In addition, we did not cover extensions of the models considered in this work. For example, the PMM may be used to include nonlinear functions of time (e.g. splines) before and/or after the changepoint. 
\\
To conclude, to our knowledge this is the first work that discusses and contrasts five mixed models capturing non-linearity of change, including LMM-quadratic, LMM-splines, FMM, PMM, SMM. This study provided insights into how each strategy can help to better characterize cognitive change over time. In addition, through a simulation study, we provided recommendations on when and how to apply these models (e.g. sample size and follow-up intervals). By performing simulations, we also emphasized that LMM-quadratic is not recommended in the context of this data. Lastly, we highlighted that more sophisticated models for longitudinal data, such as the LMM-splines, PMM, SMM, and FMM models should be considered more frequently for use in applications than is currently the case. We hope that providing details and a side-by-side comparison of what these models will encourage the adoption of these models, especially in research on cognitive aging. This paper can be seen as a support document that can facilitate the application, interpretation, and understanding of some of the complexities of the aforementioned approaches for future users.

\section*{Acknowledgments}
This work was realized on behalf of the Alzheimer’s Association International Society to Advance Alzheimer’s Research and Treatment (ISTAART), Design and Data Analytics Professional Interest Area. We thank the Catholic nuns, priests, and brothers who participated in the Religious Orders Study and participants of the Rush Memory and Aging Project. We also thank key staff members; Traci Colvin, MPH, for coordination of the clinical data collection; Karen Skish, MS, for coordination of the pathologic data collection; and John Gibbons, MS, and Greg Klein, MS, for data management. Dr. Maude Wagner is supported by a post-doctoral fellowship from the French Foundation for Alzheimer’s Research (alzheimer-recherche.org). 

\section*{Abbreviations}
\noindent BIC, Bayesian information criterion
\\FMM: functional mixed model
\\LMM, linear mixed model
\\PMM, piecewise mixed model
\\ SMM: sigmoidal mixed model
\\REML: Residual maximum likelihood
\\MSE: Mean squared error

\section*{Conflicts of interest}
The authors have no conflicts to declare. 

\section*{Fundings}
The ROS and MAP studies were funded by NIH (R01AG17917, P30AG10161, R01AG15815, R01AG34374) and the Illinois Department of Public Health. The funding organizations had no role in the design or conduct of the study; the collection, management, analysis, or interpretation of the data; or the writing of the report or the decision to submit it for publication. 

\section*{Data Availability Statement}
All ROSMAP data in these analyses (and descriptions of the studies and variables) can be requested through the Rush Alzheimer’s Disease Center Research Resource Sharing Hub at www.radc.rush.edu. The R code to replicate the analyses of this study can be provided by the corresponding author upon reasonable request.

\section*{Authors' contributions}
The authors assume full responsibility for analyses and interpretation of these data. AWC conceptualized the paper. MW performed the analysis. AWC and MW were involved in study design, interpretation of the data, drafting of the manuscript and critically revised the manuscript. DRH, GMT, and TW conducted the interpretation of the data and critically revised the manuscript. 

\section*{Competing interests}
The authors declare that they have no competing interests.

\footnotesize
\section*{References}
\noindent 1.  Steinerman JR, Hall CB, Sliwinski MJ, Lipton RB. Modeling Cognitive Trajectories Within Longitudinal Studies: A Focus on Older Adults. J Am Geriatr Soc. 2010;58(s2):S313–8. 
\\
2.	Wilson RS, Wang T, Yu L, Bennett DA, Boyle PA. Normative Cognitive Decline in Old Age. Ann Neurol. 2020;87(6):816–29. 
\\
3.	Wilson RS, Yu L, Leurgans SE, Bennett DA, Boyle PA. Proportion of cognitive loss attributable to terminal decline. Neurology. 2020;94(1):e42–50. 
\\
4.	Amieva H, Le Goff M, Millet X, Orgogozo JM, Pérès K, Barberger-Gateau P, et al. Prodromal Alzheimer’s disease: Successive emergence of the clinical symptoms. Ann Neurol. 2008;64(5):492–8. 
\\
5.	Hedeker D, Gibbons R. Longitudinal Data Analysis. Wiley Series in Probability and Statistics. John Wiley \& Sons, Hoboken. 2006. 
\\
6.	Amieva H, Mokri H, Le Goff M, Meillon C, Jacqmin-Gadda H, Foubert-Samier A, et al. Compensatory mechanisms in higher-educated subjects with Alzheimer’s disease: a study of 20 years of cognitive decline. Brain. 2014;137(4):1167–75. 
\\
7.	Terrera GM, Minett T, Brayne C, Matthews FE. Education associated with a delayed onset of terminal decline. Age Ageing. 2014;43(1):26–31. 
\\
8.	Laird, NM, Ware, JH. Random-effects models for longitudinal data. Biometrics. 1982;38(4):963–74. 
\\
9.	Cadar D, Stephan BCM, Jagger C, Johansson B, Hofer SM, Piccinin AM, et al. The role of cognitive reserve on terminal decline: a cross-cohort analysis from two European studies: OCTO-Twin, Sweden, and Newcastle 85+, UK. Int J Geriatr Psychiatry. 2016;31(6):601–10. 
\\
10.	Almkvist O, Rodriguez-Vieitez E, Thordardottir S, Nordberg A, Viitanen M, Lannfelt L, et al. Longitudinal cognitive decline in autosomal-dominant Alzheimer’s disease varies with mutations in APP and PSEN1 genes. Neurobiol Aging. 2019;82:40–7. 
\\
11.	Giil LM, Aarsland D. Greater Variability in Cognitive Decline in Lewy Body Dementia Compared to Alzheimer’s Disease. J Alzheimers Dis. 2020;73(4):1321–30. 
\\
12.	van der Willik KD, Licher S, Vinke EJ, Knol MJ, Darweesh SKL, van der Geest JN, et al. Trajectories of Cognitive and Motor Function Between Ages 45 and 90 Years: A Population-Based Study. J Gerontol Ser A. 2021;76(2):297–306. 
\\
13.	Wagner M, Wilson RS, Leurgans SE, Boyle PA, Bennett DA, Grodstein F, et al. Quantifying longitudinal cognitive resilience to Alzheimer’s disease and other neuropathologies. Alzheimers Dement. 2022;18(11):2252–61. 
\\
14.	Rice JA, Wu CO. Nonparametric Mixed Effects Models for Unequally Sampled Noisy Curves. Biometrics. 2001;57(1):253–9. 
\\
15.	Guo W. Functional Mixed Effects Models. Biometrics. 2002;58(1):121–8. 
\\
16.	Hall CB, Ying J, Kuo L, Lipton RB. Bayesian and profile likelihood change point methods for modeling cognitive function over time. Comput Stat Data Anal. 2003;42(1):91–109. 
\\
17.	Dominicus A, Ripatti S, Pedersen NL, Palmgren J. A random change point model for assessing variability in repeated measures of cognitive function. Stat Med. 2008;27(27):5786–98. 
\\
18.	van den Hout A, Muniz-Terrera G, Matthews FE. Smooth random change point models. Stat Med. 2011;30(6):599–610. 
\\
19.	Gottschalk PG, Dunn JR. The five-parameter logistic: A characterization and comparison with the four-parameter logistic. Anal Biochem. 2005;343(1):54–65. 
\\
20.	Capuano AW, Wilson RS, Leurgans SE, Dawson JD, Bennett DA, Hedeker D. Sigmoidal mixed models for longitudinal data. Stat Methods Med Res. 2018;27(3):863–75. 
\\
21. Eilers P, Marx B, Durbán M. Twenty years of P-splines. Sort-statistics and Operations Research Transactions. 2015; 39:149-86.
\\
22.	Perperoglou A, Sauerbrei W, Abrahamowicz M, Schmid M. A review of spline function procedures in R. BMC Med Res Methodol. 2019;19(1):46. 
\\
23.	Wood SN. Generalized Additive Models: An Introduction with R, Second Edition. 2nd ed. Boca Raton: Chapman and Hall/CRC; 2017. 496 p. 
\\
24.	Schwarz G. Estimating the Dimension of a Model. Ann Stat. 1978;6(2):461–4. 
\\
25.	Shi M, Weiss RE, Taylor JMG. An Analysis of Paediatric CD4 Counts for Acquired Immune Deficiency Syndrome Using Flexible Random Curves. J R Stat Soc Ser C Appl Stat. 1996;45(2):151–63. 
\\
26.	Eilers PHC, Marx BD. Flexible smoothing with B-splines and penalties. Stat Sci. 1996;11(2):89–121. 
\\
27.	HINKLEY DV. Inference about the intersection in two-phase regression. Biometrika. 1969;56(3):495–504. 
\\
28.	Hall CB, Lipton RB, Sliwinski M, Stewart WF. A change point model for estimating the onset of cognitive decline in preclinical Alzheimer’s disease. Stat Med. 2000;19(11–12):1555–66. 
\\
29.	Yu L, Boyle P, Wilson RS, Segawa E, Leurgans S, De Jager PL, et al. A Random Change Point Model for Cognitive Decline in Alzheimer’s Disease and Mild Cognitive Impairment. Neuroepidemiology. 2012;39(2):73–83. 
\\
30.	Segalas C, Amieva H, Jacqmin-Gadda H. A hypothesis testing procedure for random changepoint mixed models. Stat Med. 2019;38(20):3791–803. 
\\
31.	BACON DW, WATTS DG. Estimating the transition between two intersecting straight lines. Biometrika. 1971;58(3):525–34. 
\\
32.	Segalas C. Inférence dans les modèles à changement de pente aléatoire: application au déclin cognitif pré-démence. Bordeaux University, Bordeaux University.; 2019. 
\\
33.	PRENTICE RL. Covariate measurement errors and parameter estimation in a failure time regression model. Biometrika. 1982;69(2):331–42. 
\\
34.	Bennett DA, Buchman AS, Boyle PA, Barnes LL, Wilson RS, Schneider JA. Religious Orders Study and Rush Memory and Aging Project. J Alzheimers Dis. 2018;64(s1):S161–89. 
\\
35.	Proust-Lima C, Philipps V, Liquet B. Estimation of Extended Mixed Models Using Latent Classes and Latent Processes: The R Package lcmm. J Stat Softw. 2017;78:1–56. 
\\
36.	Wang W, Yan J. Shape-Restricted Regression Splines with R Package splines2. J Data Sci. 2021;19(3):498–517. 
\\
37.	Bates D, Mächler M, Bolker B, Walker S. Fitting Linear Mixed-Effects Models Using lme4. J Stat Softw. 2015;67:1–48. 
\\
38.	Pinheiro JC, Douglas, M, Bates. Mixed-Effects Models in S and S-PLUS. 2000. 
\\
39.	Capuano A, Wagner M. SOFTWARE nlive: an R Package to facilitate the application of the sigmoidal and random changepoint mixed models. Res Sq. 2023;rs.3.rs-2235106. 
\\
40.	Comets E, Lavenu A, Lavielle M. Parameter Estimation in Nonlinear Mixed Effect Models Using saemix, an R Implementation of the SAEM Algorithm. J Stat Softw. 2017;80:1–41. 
\\
41.	R Core Team. R: A Language and Environment for Statistical Computing. R Foundation for Statistical Computing, Vienna, Austria. 2022. 
\\
42.	Muggeo, Vito Michele Rosario and Muggeo Vm. segmented: An R package to Fit Regression Models with Broken-Line Relationships. 2008.
\\
43.	Stern Y. Cognitive reserve. Neuropsychologia. 2009;47(10):2015–28. 
\\
44.	Stern Y. Cognitive reserve in ageing and Alzheimer’s disease. Lancet Neurol. 2012;11(11):1006–12. 
\\
45.	Laukka EJ, MacDonald SWS, Bäckman L. Contrasting cognitive trajectories of impending death and preclinical dementia in the very old. Neurology. 2006;66(6):833–8. 
\\
46.	Piccinin AM, Muniz G, Matthews FE, Johansson B. Terminal Decline From Within- and Between-Person Perspectives, Accounting for Incident Dementia. J Gerontol Ser B. 2011;66B(4):391–401. 
\\
47.	Wilson RS, Yu L, Lamar M, Schneider JA, Boyle PA, Bennett DA. Education and cognitive reserve in old age. Neurology. 2019;92(10):e1041–50. 
\\
48.	Proust-Lima C, Dartigues JF, Jacqmin-Gadda H. Misuse of the Linear Mixed Model When Evaluating Risk Factors of Cognitive Decline. Am J Epidemiol. 2011;174(9):1077–88. 
\\
49.	Folstein MF, Folstein SE, McHugh PR. “Mini-mental state”. A practical method for grading the cognitive state of patients for the clinician. J Psychiatr Res. 1975;12(3):189–98. 
\\
50.	Tombaugh TN, McIntyre NJ. The Mini-Mental State Examination: A Comprehensive Review. J Am Geriatr Soc. 1992;40(9):922–35. 
\\
51.	Philipps V, Amieva H, Andrieu S, Dufouil C, Berr C, Dartigues JF, et al. Normalized Mini-Mental State Examination for Assessing Cognitive Change in Population-Based Brain Aging Studies. Neuroepidemiology. 2014;43(1):15–25.

\newpage
\section*{Appendix}

\noindent \textbf{SUPPLEMENTARY FIGURE 1.} 
Panel A) Estimated marginal trajectories of global cognition before death for LMM-splines models with K = 1, 2, 3, or 4 inner knots placed at quantiles. In this study, we selected the model with 3 inner knots, which showed a substantial gain in flexibility of the marginal trajectory compared with the model with 4 inner knots and a low Bayesian information criterion (BIC). Panel B) Representation of the four basis functions of natural cubic splines used in the approximation of the function of time in the selected LMM-splines with 3 inner knots (degrees of freedom [d] = 4)
\begin{figure}[ht!]
\includegraphics[width=450pt]{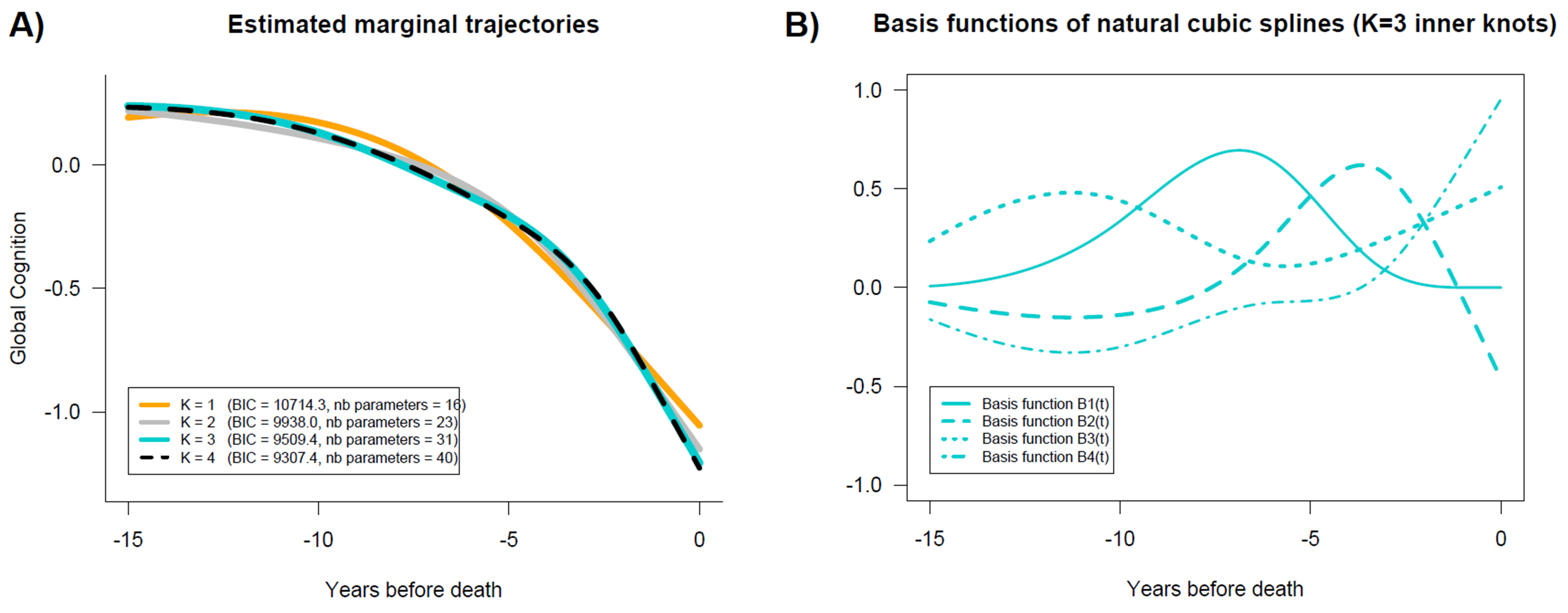}
\end{figure}

\noindent \textbf{SUPPLEMENTARY FIGURE 2.} 
Boxplots of the longitudinal observed measures, obtained every year before death, from the simulated data in Scenario A and B (n=1000)
\begin{figure}[ht!]
\includegraphics[width=460pt]{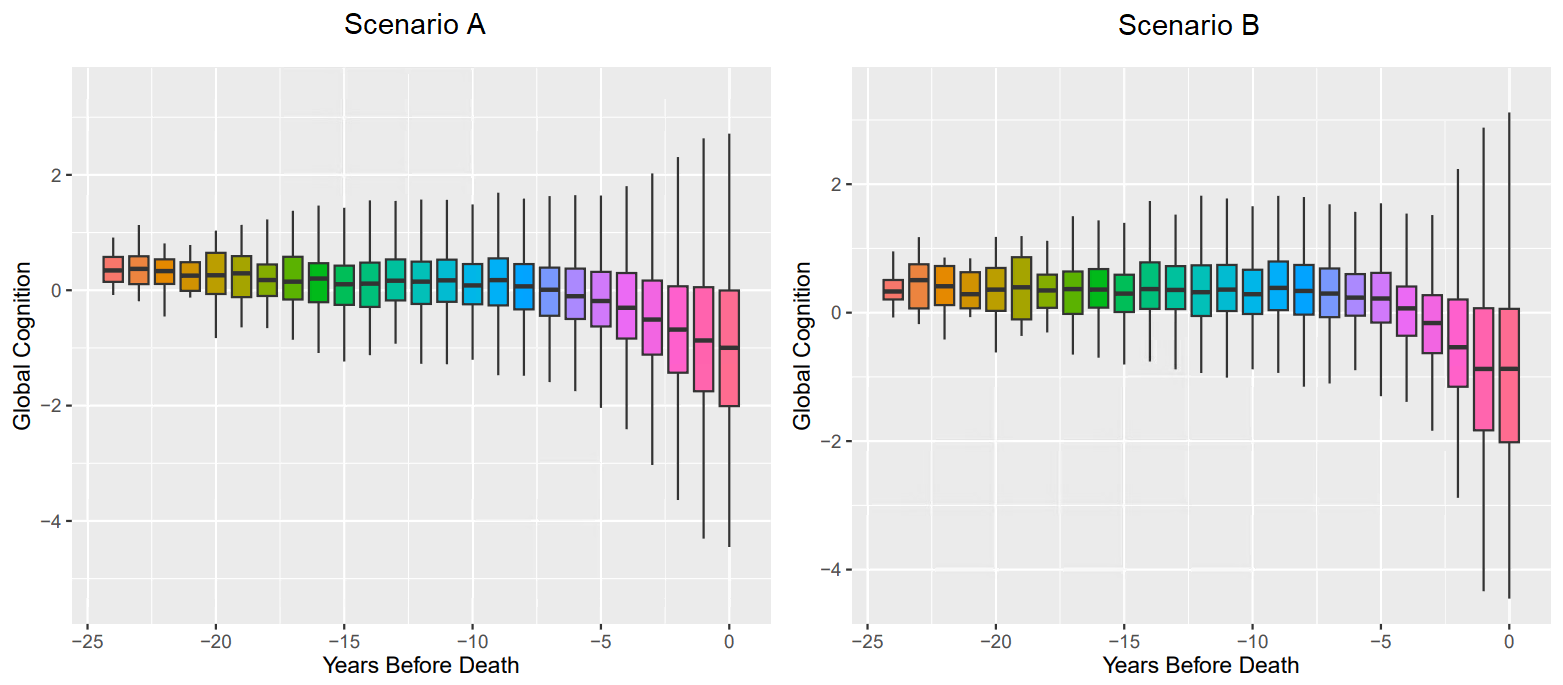}
\end{figure}

\begin{landscape}
\noindent \textbf{SUPPLEMENTARY TABLE 1.} Mean Squared Errors (MSE) and Bias from the average of the marginal trajectories of global cognition estimated in the 15 years before death across 500 simulations of 1,000 individuals each, for the five models (Scenarios A and B). LMM, linear mixed model; PMM, piecewise mixed model; SMM: sigmoidal mixed model; FMM: functional mixed model.
\begin{figure}[ht!]
\includegraphics[width=630pt]{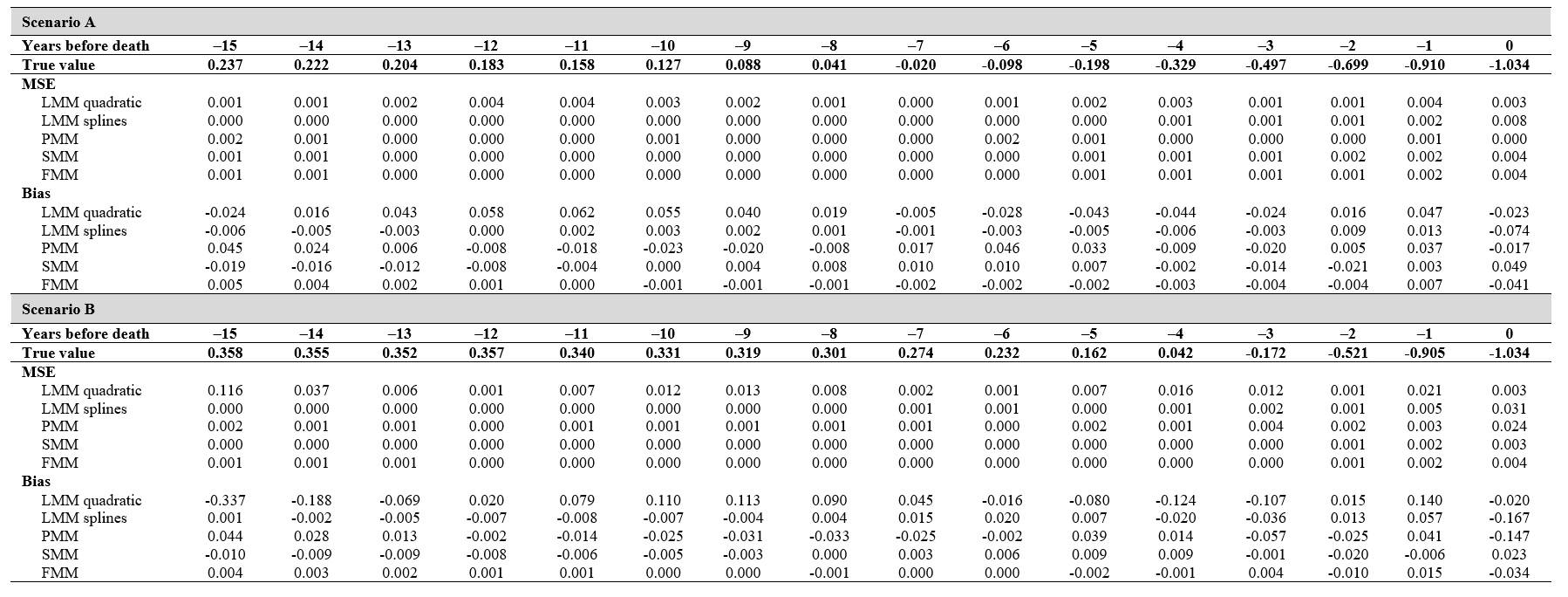}
\end{figure}
\end{landscape}

\noindent \textbf{SUPPLEMENTARY FIGURE 3.} 
Averaged marginal trajectories of global cognition estimated before death across 500 simulations of 1,000 individuals each, using the five models and after excluding the last cognitive assessment prior to death for 30\% of individuals for Scenario A1 and Scenario B1. LMM, linear mixed model; PMM, piecewise mixed model; SMM: sigmoidal mixed model; FMM: functional mixed model
\begin{figure}[ht!]
\includegraphics[width=420pt]{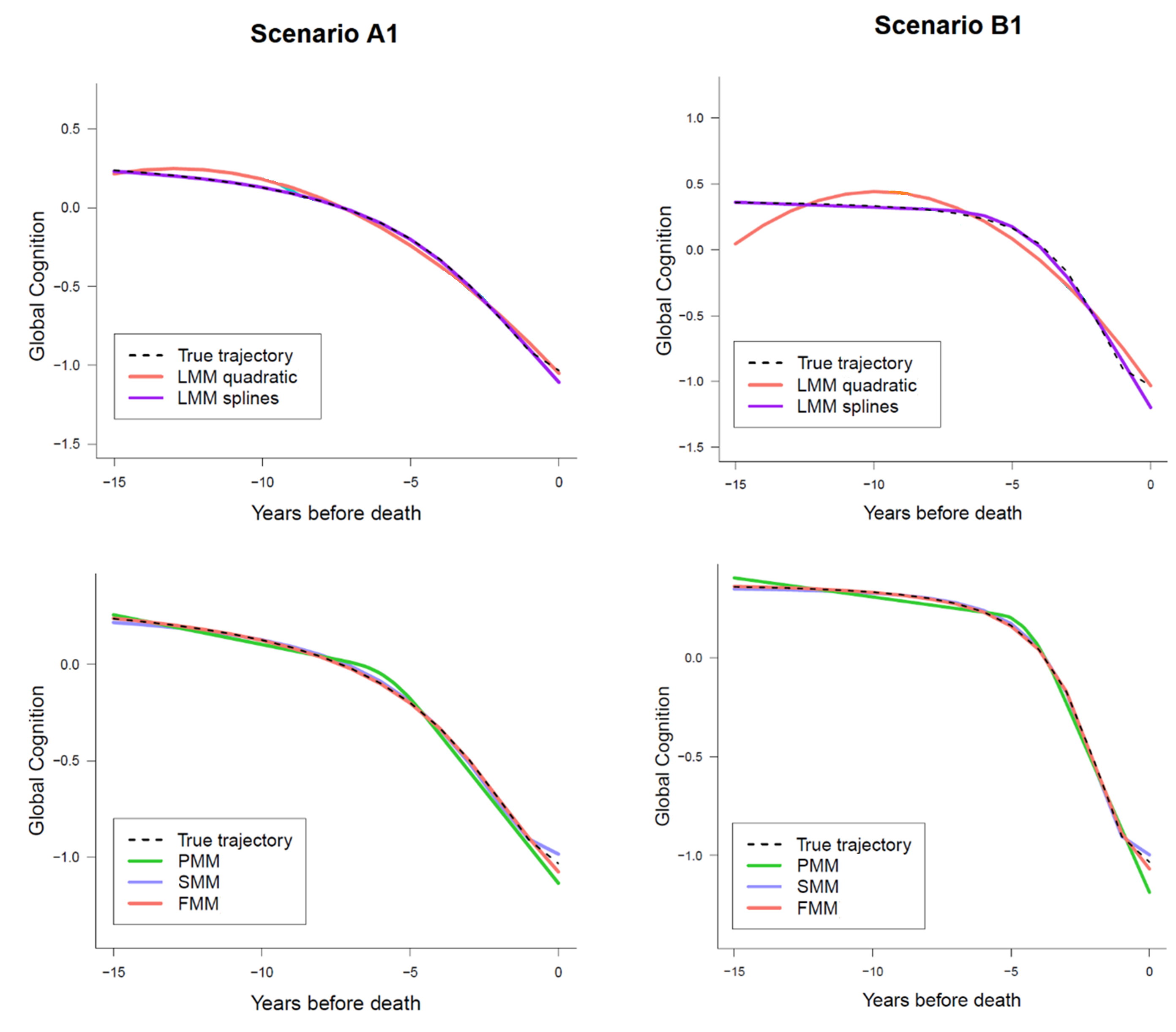}
\end{figure}

\begin{landscape}
\noindent \textbf{SUPPLEMENTARY TABLE 2.} Mean Squared Errors (MSE) and bias from the average of the marginal trajectories of global cognition estimated in the 15 years before death across 500 simulations of 1,000 individuals each, for the five models after excluding the last cognitive assessment prior to death for 30\% of individuals (Scenarios A1 and B1). LMM, linear mixed model; PMM, piecewise mixed model; SMM: sigmoidal mixed model; FMM: functional mixed model.
\begin{figure}[ht!]
\includegraphics[width=570pt]{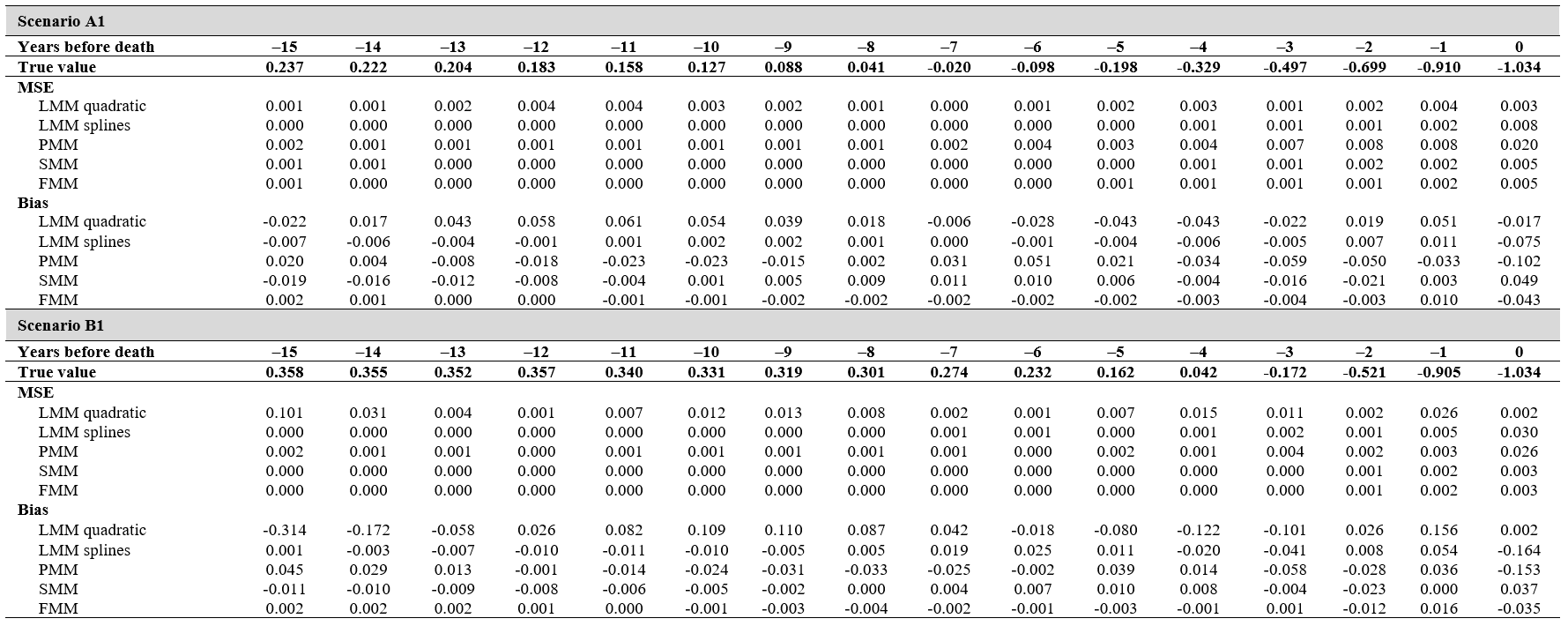}
\end{figure}
\end{landscape}

\noindent \textbf{SUPPLEMENTARY FIGURE 4.} 
Averaged marginal trajectories of global cognition estimated before death across 500 simulations of 1,000 individuals each, using the five models and when considering that half of individuals (n=500) had only 4 cognitive measures prior to death for Scenario A2 and Scenario B2. LMM, linear mixed model; PMM, piecewise mixed model; SMM: sigmoidal mixed model; FMM: functional mixed model
\begin{figure}[ht!]
\includegraphics[width=420pt]{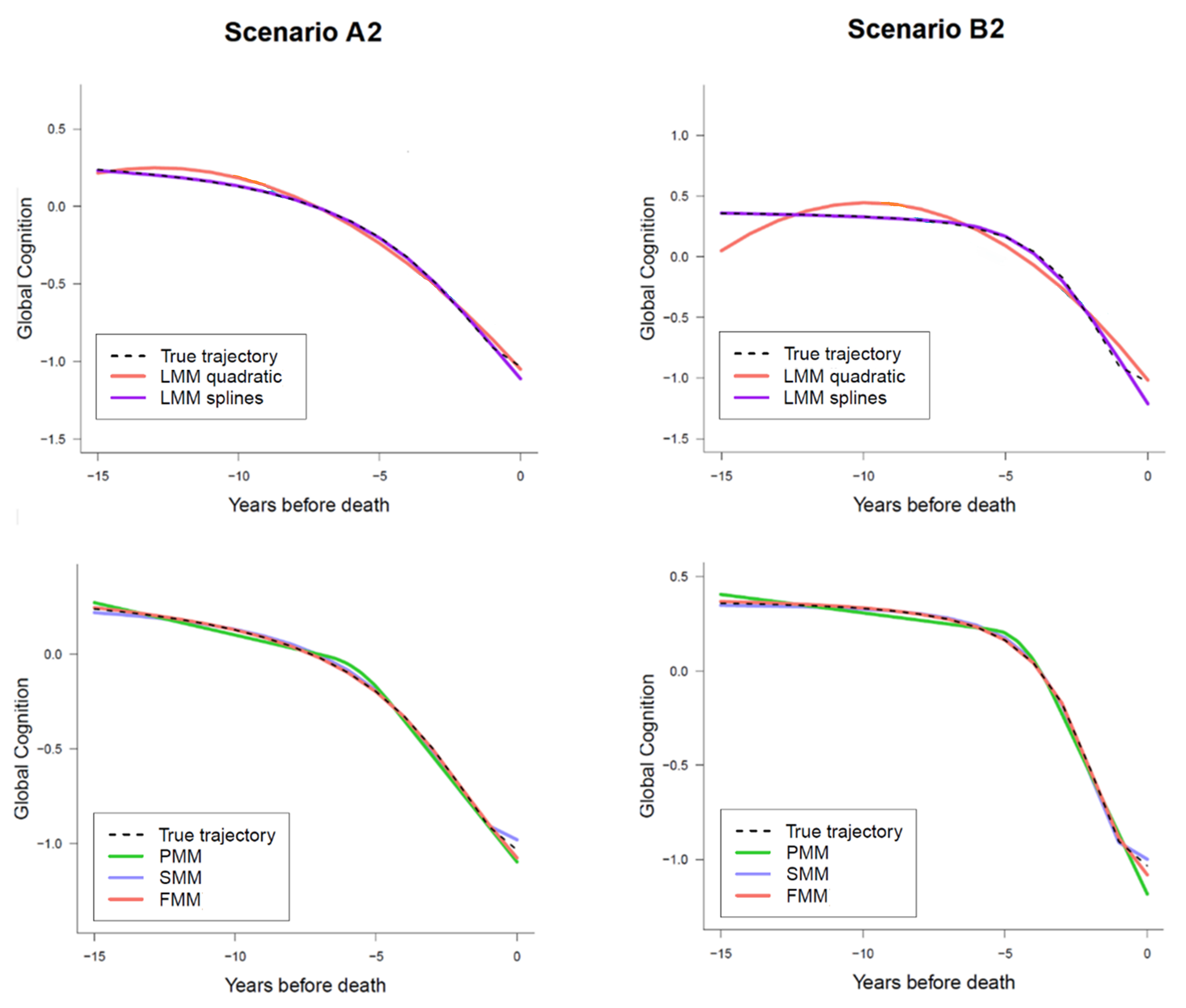}
\end{figure}

\begin{landscape}
\noindent \textbf{SUPPLEMENTARY TABLE 3.} Mean Squared Errors (MSE) and Bias from the average of the marginal trajectories of global cognition estimated in the 15 years before death across 500 simulations of 1,000 individuals each, for the five models and when considering that half of individuals (n=500) had only 4 cognitive measures prior to death (Scenarios A2 and B2). LMM, linear mixed model; PMM, piecewise mixed model; SMM: sigmoidal mixed model; FMM: functional mixed model.
\begin{figure}[ht!]
\includegraphics[width=570pt]{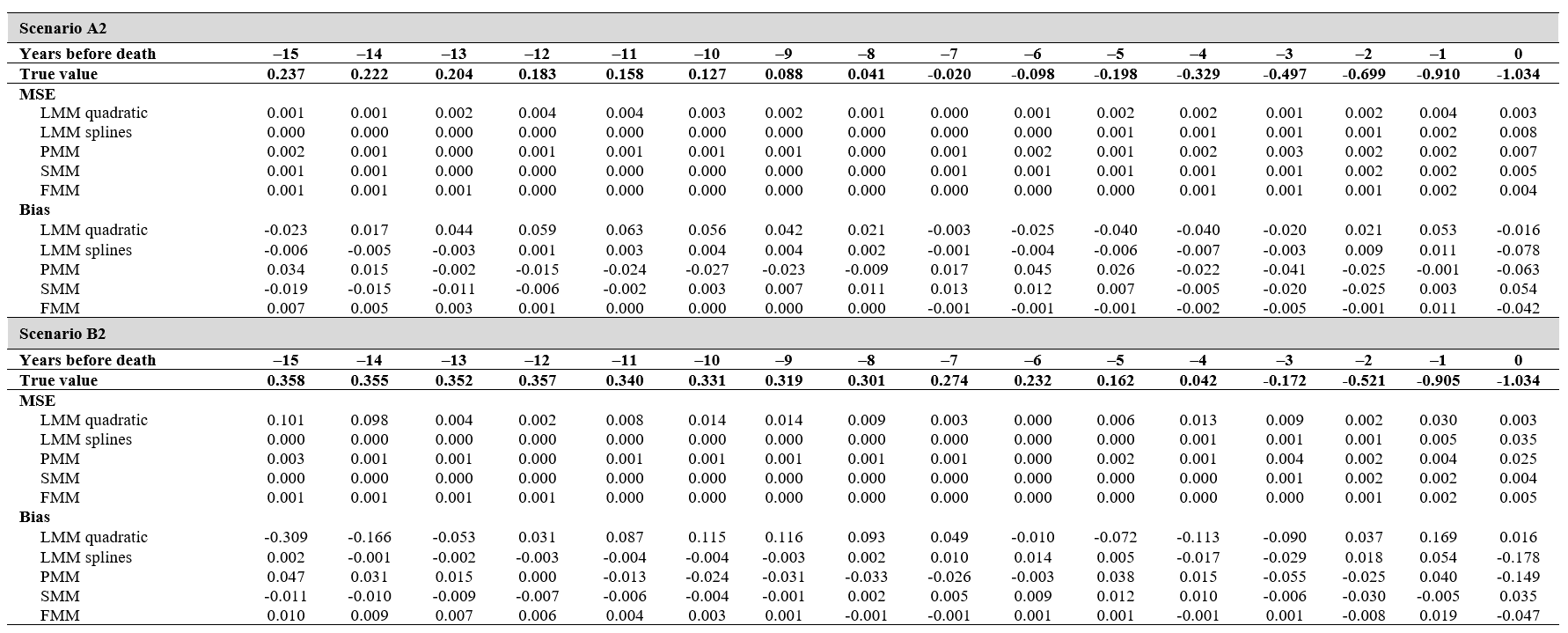}
\end{figure}
\end{landscape}

\noindent \textbf{SUPPLEMENTARY FIGURE 5.} 
Averaged marginal trajectories of global cognition estimated before death across 500 simulations of 1,000 individuals each, using the five models and when considering cognitive assessment every 3 years instead of every year for Scenario A3 and Scenario B3. LMM, linear mixed model; PMM, piecewise mixed model; SMM: sigmoidal mixed model; FMM: functional mixed model
\begin{figure}[ht!]
\includegraphics[width=420pt]{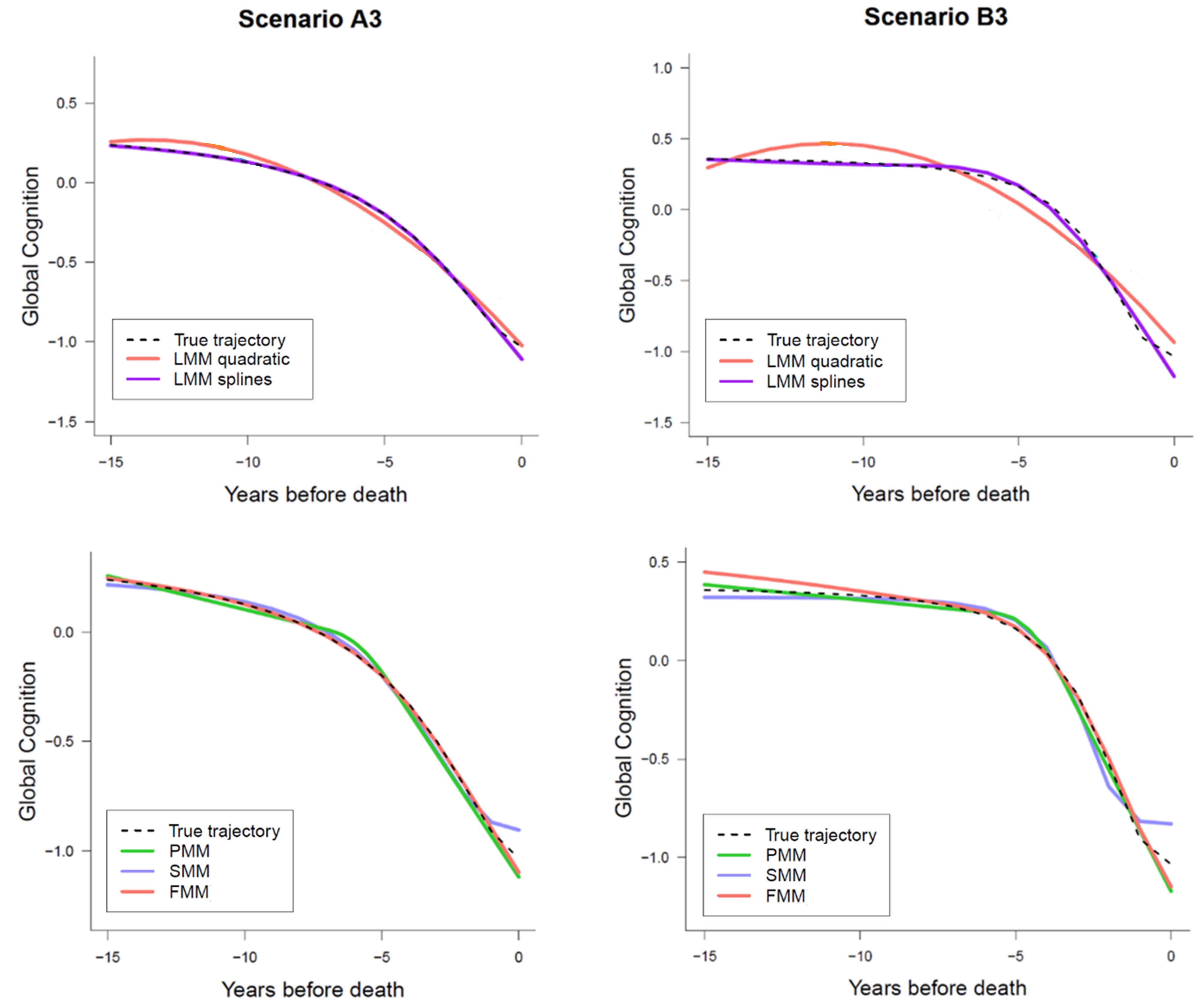}
\end{figure}

\begin{landscape}
\noindent \textbf{SUPPLEMENTARY TABLE 4.} Mean Squared Errors (MSE) and Bias from the average of the marginal trajectories of global cognition estimated in the 15 years before death across 500 simulations of 1,000 individuals each, for the five models and when considering cognitive assessment every 3 years instead of every year (Scenarios A3 and B3). LMM, linear mixed model; PMM, piecewise mixed model; SMM: sigmoidal mixed model; FMM: functional mixed model.
\begin{figure}[ht!]
\includegraphics[width=570pt]{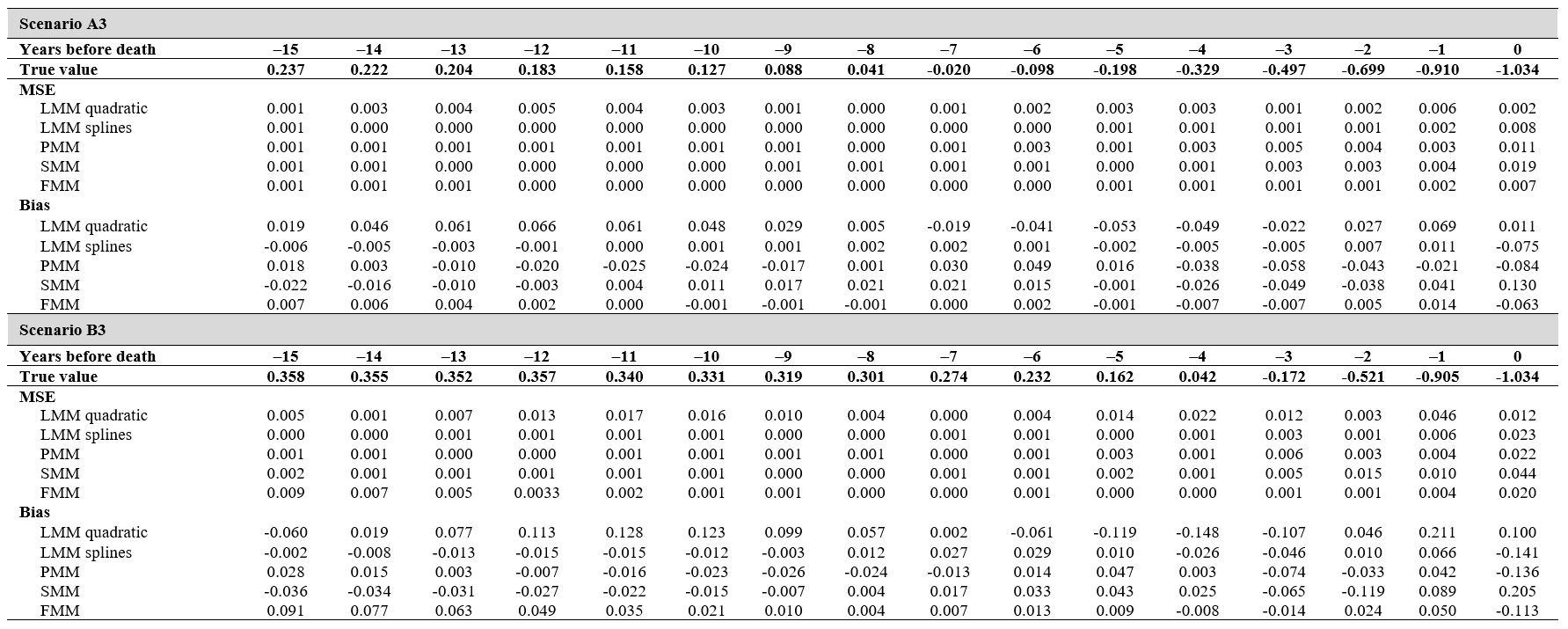}
\end{figure}
\end{landscape}

\noindent \textbf{SUPPLEMENTARY FIGURE 6.} 
Averaged marginal trajectories of global cognition estimated before death across 500 simulations of 150 individuals each, using the five models for Scenario A4 and Scenario B4. LMM, linear mixed model; PMM, piecewise mixed model; SMM: sigmoidal mixed model; FMM: functional mixed model
\begin{figure}[ht!]
\includegraphics[width=420pt]{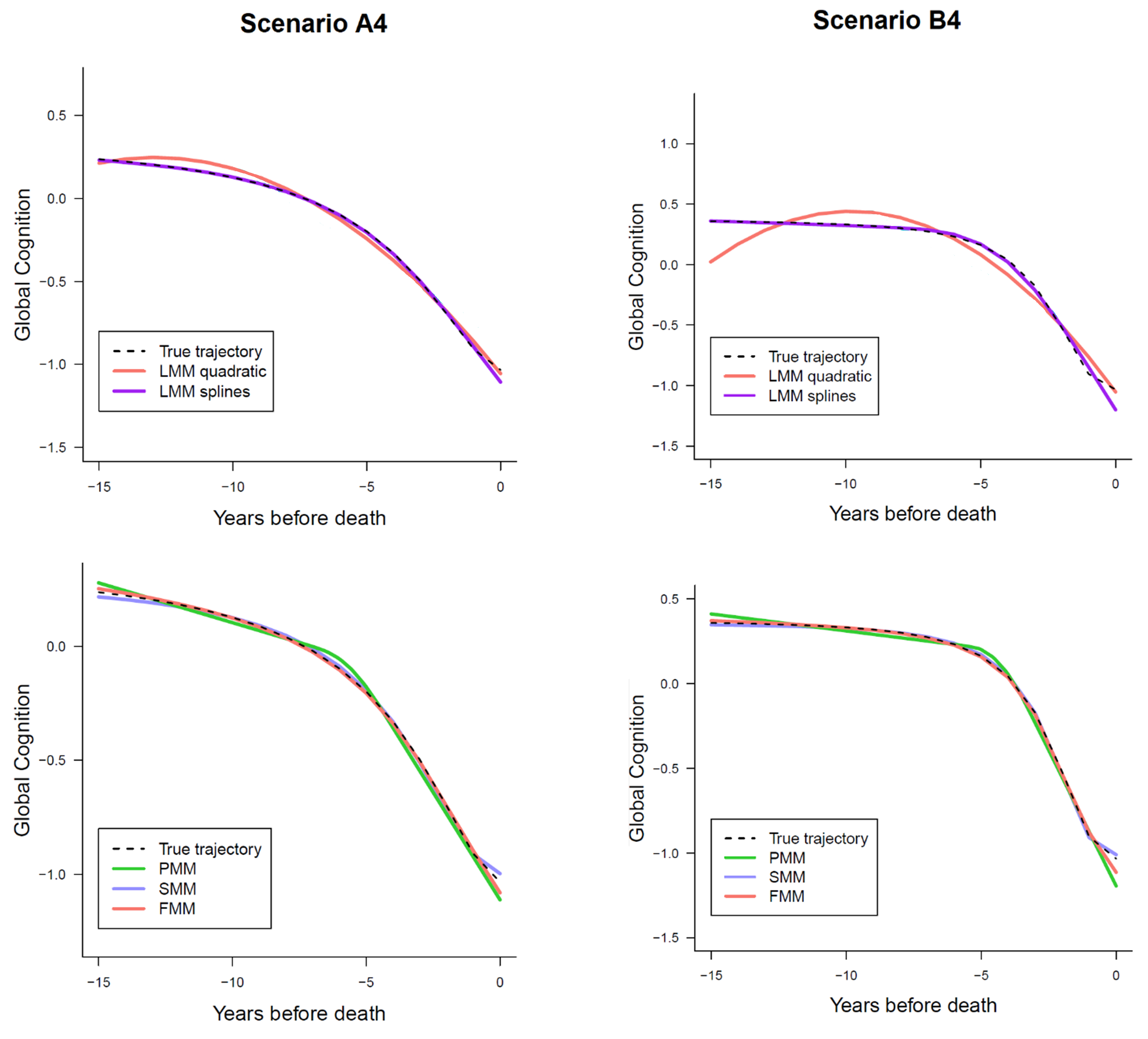}
\end{figure}

\begin{landscape}
\noindent \textbf{SUPPLEMENTARY TABLE 5.} Mean Squared Errors (MSE) and Bias from the average of the marginal trajectories of global cognition estimated in the 15 years before death across 500 simulations of 150 individuals each, for the five models (Scenarios A4 and B4). LMM, linear mixed model; PMM, piecewise mixed model; SMM: sigmoidal mixed model; FMM: functional mixed model.
\begin{figure}[ht!]
\includegraphics[width=570pt]{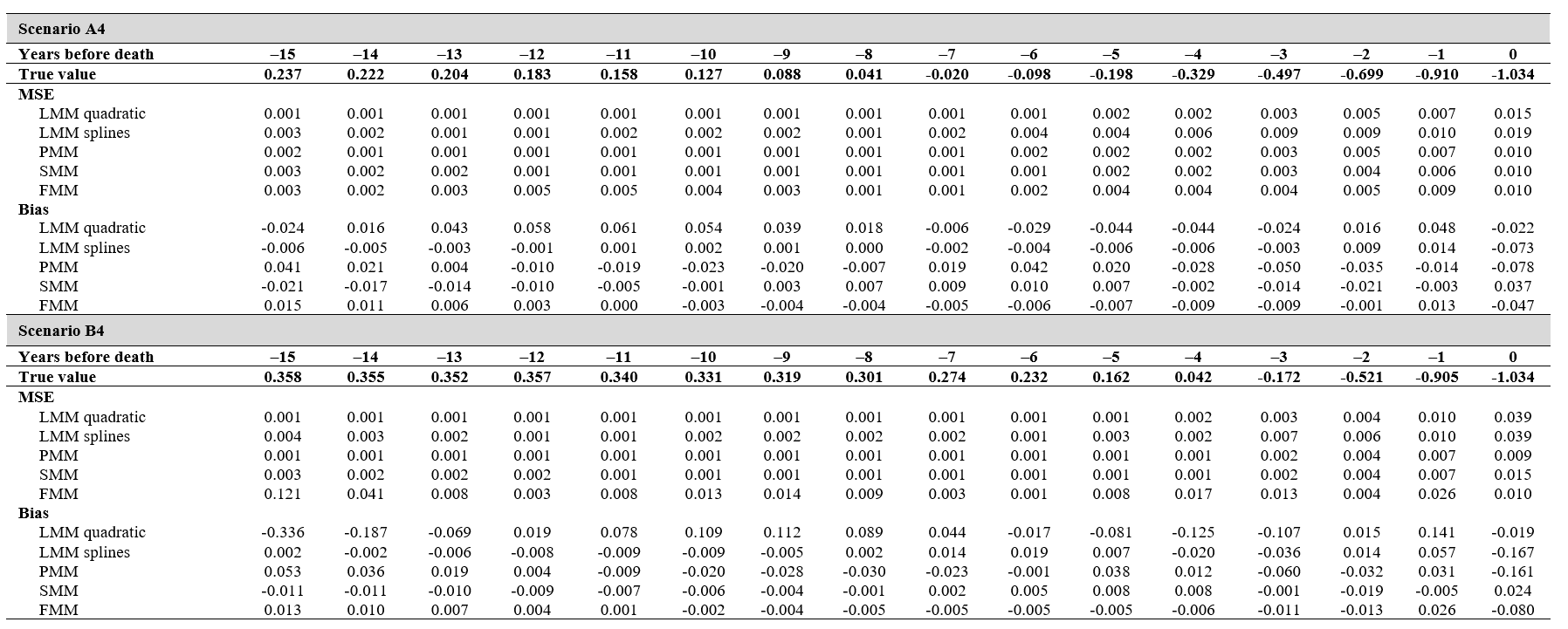}
\end{figure}
\end{landscape}


\begin{thebibliography}{10}

\bibitem{Hirt1974}
Hirt CW, Amsden AA, Cook JL. An arbitrary {L}agrangian-{E}ulerian computing
  method for all flow speeds.  {\it J {C}omput {P}hys. }1974;14(3):227--253.

\bibitem{Liska2010}
Liska R, Shashkov M, Vachal P, Wendroff B. Optimization-based synchronized
  flux-corrected conservative interpolation (remapping) of mass and momentum
  for arbitrary {L}agrangian-{E}ulerian methods.  {\it J {C}omput {P}hys.
  }2010;229(5):1467--1497.

\bibitem{Taylor1937}
Taylor GI, Green AE. Mechanism of the production of small eddies from large
  ones.  {\it P {R}oy {S}oc {L}ond {A} {M}at. }1937;158(895):499--521.
\newblock \url{https://doi.org/10.1098/rspa.1937.0036},
  \url{http://rspa.royalsocietypublishing.org/content/158/895/499}.

\bibitem{Knupp1999}
Knupp PM. Winslow smoothing on two-dimensional unstructured meshes.  {\it Eng
  {C}omput. }1999;15:263--268.

\bibitem{Kamm2000}
Kamm J. {\it Evaluation of the {S}edov-von {N}eumann-{T}aylor blast wave
  solution. } Technical {R}eport LA-UR-00-6055: Los {A}lamos {N}ational
  {L}aboratory; 2000.

\bibitem{Kucharik2003}
Kucharik M, Shashkov M, Wendroff B. An efficient linearity-and-bound-preserving
  remapping method.  {\it J {C}omput {P}hys. }2003;188(2):462--471.

\bibitem{Blanchard2015}
Blanchard G, Loubere R. {\it High-Order {C}onservative {R}emapping with a
  posteriori {MOOD} stabilization on polygonal meshes. }
  \url{https://hal.archives-ouvertes.fr/hal-01207156}, the {HAL} {O}pen
  {A}rchive, hal-01207156. Accessed January 13, 2016; 2015.

\bibitem{Burton2013}
Burton DE, Kenamond MA, Morgan NR, Carney TC, Shashkov MJ. An intersection
  based {ALE} scheme {(xALE)} for cell centered hydrodynamics {(CCH)}.  In:
  Talk at {M}ultimat 2013, {I}nternational {C}onference on {N}umerical
  {M}ethods for {M}ulti-{M}aterial {F}luid {F}lows; September 2--6, 2013; San
  {F}rancisco.
\newblock LA-UR-13-26756.2.

\bibitem{Berndt2011}
Berndt M, Breil J, Galera S, Kucharik M, Maire PH, Shashkov M. Two-step hybrid
  conservative remapping for multimaterial arbitrary {L}agrangian-{E}ulerian
  methods.  {\it J {C}omput {P}hys. }2011;230(17):6664--6687.

\bibitem{Kucharik2012}
Kucharik M, Shashkov M. One-step hybrid remapping algorithm for multi-material
  arbitrary {L}agrangian-{E}ulerian methods.  {\it J {C}omput {P}hys.
  }2012;231(7):2851--2864.

\bibitem{Breil2015}
Breil J, Alcin H, Maire PH. A swept intersection-based remapping method for
  axisymmetric {ReALE} computation.  {\it Int {J} {N}umer {M}eth {F}l.
  }2015;77(11):694--706.
\newblock Fld.3996.

\bibitem{Barth1997}
Barth TJ. Numerical methods for gasdynamic systems on unstructured meshes.  In:
   Kroner D, Rohde C, Ohlberger M, eds. {\it An {I}ntroduction to {R}ecent
  {D}evelopments in {T}heory and {N}umerics for {C}onservation {L}aws,
  {P}roceedings of the {I}nternational {S}chool on {T}heory and {N}umerics for
  {C}onservation {L}aws}, Lecture {N}otes in {C}omputational {S}cience and
  {E}ngineering. Berlin: Springer 1997.
\newblock ISBN 3-540-65081-4.

\bibitem{Lauritzen2011}
Lauritzen P, Erath C, Mittal R. On simplifying `incremental remap'-based
  transport schemes.  {\it J {C}omput {P}hys. }2011;230(22):7957--7963.

\bibitem{Klima2017}
Klima M, Kucharik M, Shashkov M. Local error analysis and comparison of the
  swept- and intersection-based remapping methods.  {\it Commun {C}omput
  {P}hys. }2017;21(2):526--558.

\bibitem{Dukowicz2000}
Dukowicz JK, Baumgardner JR. Incremental remapping as a transport/advection
  algorithm.  {\it J {C}omput {P}hys. }2000;160(1):318--335.

\bibitem{Kucharik2011}
Kucharik M, Shashkov M. Flux-based approach for conservative remap of
  multi-material quantities in {2D} arbitrary {L}agrangian-{E}ulerian
  simulations.  In:  Fo\v{r}t J, F{\"{u}}rst J, Halama J, Herbin R, Hubert F,
  eds. {\it Finite {V}olumes for {C}omplex {A}pplications {VI} {P}roblems \&
  {P}erspectives},  Springer {P}roceedings in {M}athematics, vol. 1: Springer
  2011 (pp. 623--631).

\bibitem{Kucharik2014}
Kucharik M, Shashkov M. Conservative multi-material remap for staggered
  multi-material arbitrary {L}agrangian-{E}ulerian methods.  {\it J {C}omput
  {P}hys. }2014;258:268--304.

\bibitem{Loubere2005}
Loubere R, Shashkov M. A subcell remapping method on staggered polygonal grids
  for arbitrary-{L}agrangian-{E}ulerian methods.  {\it J {C}omput {P}hys.
  }2005;209(1):105--138.

\bibitem{Caramana1998}
Caramana EJ, Shashkov MJ. Elimination of artificial grid distortion and
  hourglass-type motions by means of {L}agrangian subzonal masses and
  pressures.  {\it J {C}omput {P}hys. }1998;142(2):521--561.

\bibitem{Hoch2009}
Hoch P. {\it An arbitrary {L}agrangian-{E}ulerian strategy to solve
  compressible fluid flows. } Technical {R}eport: CEA; 2009.
\newblock HAL: hal-00366858.
  https://hal.archives-ouvertes.fr/docs/00/36/68/58/PDF/ale2d.pdf. Accessed
  January 13, 2016.

\bibitem{Shashkov1996}
Shashkov M. {\it Conservative {F}inite-{D}ifference {M}ethods on {G}eneral
  {G}rids}.
\newblock Boca Raton, Florida: CRC {P}ress; 1996.
\newblock ISBN 0-8493-7375-1.

\bibitem{Benson1992}
Benson DJ. Computational methods in {L}agrangian and {E}ulerian hydrocodes.
  {\it Comput {M}ethod {A}ppl {M}. }1992;99(2--3):235--394.

\bibitem{Margolin2003}
Margolin LG, Shashkov M. Second-order sign-preserving conservative
  interpolation (remapping) on general grids.  {\it J {C}omput {P}hys.
  }2003;184(1):266--298.

\bibitem{Kenamond2013}
Kenamond MA, Burton DE. Exact intersection remapping of multi-material
  domain-decomposed polygonal meshes.  In: Talk at {M}ultimat 2013,
  {I}nternational {C}onference on {N}umerical {M}ethods for {M}ulti-{M}aterial
  {F}luid {F}lows; September 2--6, 2013; San {F}rancisco.
\newblock LA-UR-13-26794.

\bibitem{Dukowicz1984}
Dukowicz J. Conservative rezoning (remapping) for general quadrilateral meshes.
   {\it J {C}omput {P}hys. }1984;54(3):411--424.

\bibitem{Margolin2002}
Margolin LG, Shashkov M. {\it Second-order sign-preserving remapping on general
  grids. } Technical Report LA-UR-02-525: Los {A}lamos {N}ational {L}aboratory;
  2002.

\bibitem{Mavriplis2003}
Mavriplis DJ. Revisiting the least-squares procedure for gradient
  reconstruction on unstructured meshes.  In: AIAA 2003-3986. 16th {AIAA}
  {C}omputational {F}luid {D}ynamics {C}onference; June 23--26, 2003; Orlando,
  {F}lorida.

\bibitem{Scovazzi2008}
Scovazzi G, Love E, Shashkov M. Multi-scale {L}agrangian shock hydrodynamics on
  {Q1/P0} finite elements: {T}heoretical framework and two-dimensional
  computations.  {\it Comput {M}ethod {A}ppl {M}. }2008;197(9--12):1056--1079.

\end{thebibliography}
\end{document}